\pdfoutput=1

\documentclass[fleqn,usenatbib,useAMS]{mnras}

\usepackage{graphicx}	% Including figure files
\usepackage{amsmath}	% Advanced maths commands
\usepackage{multicol}        % Multi-column entries in tables
\usepackage{bm}		% Bold maths symbols, including upright Greek
\usepackage{pdflscape}	% Landscape pages
\usepackage{latexsym}
\usepackage[authoryear]{natbib}
\usepackage{bm,nicefrac}

\usepackage[T1]{fontenc}
\usepackage{ae,aecompl}

\usepackage{newtxtext,newtxmath}
\def\deg{\ifmmode^\circ\else$^\circ$\fi}

\def\Q{\ifmmode\mathcal{Q}\else$\mathcal{Q}$\fi}
\def\Mach{\ifmmode\mathcal{M}\else$\mathcal{M}$\fi}

\title[Ionized filaments around {\it l} = 345.5 degree]
{Ionized filaments and ongoing physical processes in massive star-forming sites around {\it l} = 345.5 degree}
\author[L.~K. Dewangan et al.]
{L.~K. Dewangan$^{1}$\thanks{lokeshd@prl.res.in}, L.~E. Pirogov$^{2}$, N.~K. Bhadari$^{1,3}$, and A.~K. Maity$^{1,3}$\\ 
$^{1}$Physical Research Laboratory, Navrangpura, Ahmedabad - 380 009, India.\\
$^{2}$Institute of Applied Physics of the Russian Academy of Sciences, 46 Ulyanov st., Nizhny Novgorod 603950, Russia.\\
$^{3}$Indian Institute of Technology Gandhinagar Palaj, Gandhinagar 382355, India.} 
\begin{document}

\date{ }

\pagerange{\pageref{firstpage}--\pageref{lastpage}} \pubyear{2020}

\maketitle

\label{firstpage}

\begin{abstract}
Numerous research studies on dust and molecular filaments have been conducted in star-forming sites, but
only a limited number of studies have focused on ionized filaments. 
To observationally study this aspect, we present an analysis of multi-wavelength data of an area of $\sim$74\rlap.{$'$}6 $\times$ 55$'$ around {\it l} = 345$\degr$.5. Using the 843 MHz continuum map, two distinct ionized filaments (i.e., IF-A (extent $\sim$8\rlap.{$'$}5) and IF-B (extent $\sim$22\rlap.{$'$}65)) hosting ionized clumps powered by massive OB stars are identified. Using the $^{13}$CO(2--1) and C$^{18}$O(2--1) line data, the parent molecular clouds of IF-A and IF-B are studied in a velocity range of [$-$21, $-$10] km s$^{-1}$, and have filamentary appearances. At least two cloud components around $-$18 and $-$15 km s$^{-1}$ toward the parent clouds of IF-A and IF-B are investigated, and are connected in velocity space. These filamentary clouds also spatially overlap with each other along the major axis, backing the filamentary twisting/coupling nature. Noticeable Class~I protostars and massive stars appear to be observed toward the common zones of the cloud components. These findings support the collision of two filamentary clouds around 1.2 Myr ago. The existence of the ionized filaments seems to be explained by the combined feedback of massive stars. The molecular filaments associated with IF-A and IF-B favour the outcomes of the most recent model concerning the escape and the trap of the ionizing radiation from an O star formed in a filament.
\end{abstract}
%------------------
%
\section{Introduction}
\label{sec:intro}
In recent years, sub-millimeter(mm) continuum and molecular-line studies have revealed molecular and dust filaments as common features in massive star-forming regions \citep[e.g.,][]{andre10,andre14,morales19}. 
The involvement of these filaments in the study of the origin of massive OB-stars (M $\gtrsim$ 8 M$_{\odot}$) has received great impetus in recent years. 
In other words, multi-scale and multi-wavelength studies of the filaments are a reliable approach to deepen understanding of massive star formation (MSF) mechanisms. It has been thought that OB stars are assembled by large-scale (1--10 pc) inflow material that may be funneled along filaments \citep[e.g.,][]{tan14,Motte+2018,hirota18,rosen20}.  
Such process favours the convergence of filaments toward the compact and dense hub, or a 
star-forming clump surrounded by filaments or a junction of 
filaments \citep[i.e., hub-filament system (HFS);][]{myers09,Motte+2018}. 
A hub-filament configuration is almost universally detected in massive star-forming regions.
Furthermore, the intersection/merging/collision of filaments can also explain the formation of 
massive OB stars and stellar clusters \citep[e.g.,][and references therein]{habe92,anathpindika10,fukui21}. Hence, multiple physical processes are expected to be operated in massive star-forming regions.

Apart from the molecular and dust filaments, one may also expect elongated filaments of ionized gas in 
star-forming regions, but such study is very limited in the literature (e.g., LBN 140.07+01.64 \citep{karr03,dewangan21}; Eridanus filaments \citep{pon14}; Cygnus~X \citep{emig22}). Hence, the simultaneous study of the ionized, dust, and molecular filaments is still lacking 
due to scarcity of the ionized filaments in star-forming regions \citep[e.g., LBN 140.07+01.64;][]{dewangan21}. 
The ionizing radiation from OB-association (or OB-star complex) is thought to be responsible for the origin of ionized filaments. Such ionized filaments are found at far distances from the exciting stars/complex \citep[e.g.,][]{karr03,pon14,emig22}. 
On the other hand, there is a possibility that several massive stars formed in molecular/dust filamentary clouds may locally produce the ionized filaments in the same clouds. However, such a proposal is yet to be explored in star-forming sites. 
Hence, such targets offer to study not only the birth of massive OB stars, but also the origin of elongated ionized filaments. 
It also enables us to study the role of filaments in MSF activities and the impact 
of massive OB stars on their parent filaments. 

In this context, the present paper deals with a wide target area around {\it l} = 345$\degr$.5, which contains several star-forming sites (e.g., IRAS 17008-4040, IRAS 17009-4042, IRAS 17006-4037, IRAS 17028-4050, IRAS 17027-4100, IRAS 17026-4053, and IRAS 17024-4106). Among these highlighted sources, IRAS 17008-4040 and IRAS 17009-4042 are well known massive star-forming regions. The selected target area is not very distant ($<$ 2.5 kpc), and hosts previously known H\,{\sc ii} regions powered by massive OB stars, dust filaments, clusters of young protostars, and a massive protostar in a young, pre-ultracompact H\,{\sc ii} phase. The selected sources are the potential targets to explore the role of filaments in MSF processes and the impact of massive stars on the filaments. Furthermore, such targets also seem to be very promising for investigating the ionized filaments and their molecular environments, which is a very poorly studied topic in star formation research.

Situated at a distance of $\sim$2.4 kpc, the sites IRAS 17008-4040 (hereafter i17008) and IRAS 17009-4042 (hereafter i17009) are associated with H\,{\sc ii} regions powered by B-type stars \citep{garay06,garay07,dewangan18}. Radio continuum morphologies of the H\,{\sc ii} regions at different radio frequencies (i.e., 0.61, 1.28, 1.4, and 2.5 GHz) were examined by \citet{dewangan18} (see Figure~9 in their paper). 
An elongated filament hosting the sites i17008 and i17009 was reported using the APEX Telescope Large Area Survey of the Galaxy \citep[ATLASGAL;][]{schuller09} 870 $\mu$m continuum map \citep[see also][]{dewangan18}. 
With the aid of the 870 $\mu$m continuum data, at least one dust continuum clump is detected toward these two IRAS sites. 
Clumps clm1 (M$_\mathrm{clump}$ $\sim$2430 M$_{\odot}$; $T_\mathrm{d}$ $\sim$30 K; V$_\mathrm{lsr}$ = $-$17 km s$^{-1}$; d $\sim$2.4 kpc) 
and clm2 (M$_\mathrm{clump}$ $\sim$2900 M$_{\odot}$; $T_\mathrm{d}$ $\sim$27.3 K; V$_\mathrm{lsr}$ = $-$17.3 km s$^{-1}$; d $\sim$2.4 kpc) are detected toward i17008 and i17009, respectively \citep[see][for more details]{urquhart18,dewangan18}. 
The study of the {\it Herschel} sub-mm maps revealed the existence of several parsec-scale filaments directed toward the dust clump hosting each IRAS site, exhibiting HFS candidates \citep{dewangan18}. Additionally, the site i17008 hosts an infrared counterpart (IRc) of the 6.7 GHz methanol maser emission (MME), which is also associated with an extended green object \citep[EGO;][]{cyganowski08}. 
The IRc has been proposed as an O-star candidate without an H\,{\sc ii} region \citep[see Figure~9 in][]{dewangan18}, which drives an outflow \citep{cyganowski08,morales09,dewangan18}. Overall, the ongoing star formation activities (including massive stars) were reported toward i17008 and i17009 using the infrared photometric data and radio continuum maps \citep{dewangan18}. 

Figures~\ref{fig1}a and~\ref{fig1}b present the radio continuum emission contours 
at 843 MHz from the Sydney University Molonglo Sky Survey \citep[SUMSS;][]{bock99} and the {\it Spitzer} Galactic Legacy Infrared Mid-Plane Survey Extraordinaire \citep[GLIMPSE;][]{benjamin03} 8.0 $\mu$m image of a wide area (size $\sim$74\rlap.{$'$}64 $\times$ 55\rlap.{$'$}02; centered at {\it l} = 345$\degr$.3693; {\it b} = 0$\degr$.0391), respectively, which is the target area of this paper.  
At least two elongated morphologies or ionized filaments appear in the 843 MHz continuum map toward our selected area around {\it l} = 345$\degr$.5 (see two dotted-dashed boxes in Figure~\ref{fig1}a), where the extended emission in the 8.0 $\mu$m image is also traced 
(see Figure~\ref{fig1}b and also Section~\ref{sec:morph} for more details). 
We do not find any study of these elongated ionized filaments in the literature as well as their association with the 
dust and molecular filaments. 

There is no understanding of the existence of these structures and of the ongoing physical mechanisms 
around {\it l} = 345$\degr$.5. 
In this context, to observationally study the formation of massive stars and the origin of the ionized filaments, an extensive analysis of the multi-wavelength data sets (see Section~\ref{sec:obser}) is performed. 
In particular, to study the parent molecular clouds of the ionized filaments, we analyzed the unexplored molecular line data from the structure, excitation, and dynamics of the inner Galactic interstellar medium \citep[SEDIGISM;][]{schuller17,schuller21} and the Mopra Southern Galactic Plane CO Survey \citep{braiding18}. 

Section~\ref{sec:obser} presents the observational data sets discussed in this paper. 
The outcomes of this paper are given in Section~\ref{sec:data}. 
In Section~\ref{sec:disc}, the implications of our observed outcomes are discussed. 
Finally, Section~\ref{sec:conc} gives the conclusions of this study.
\section{Data sets}
\label{sec:obser}
\begin{figure}
\center
\includegraphics[width=8.5cm]{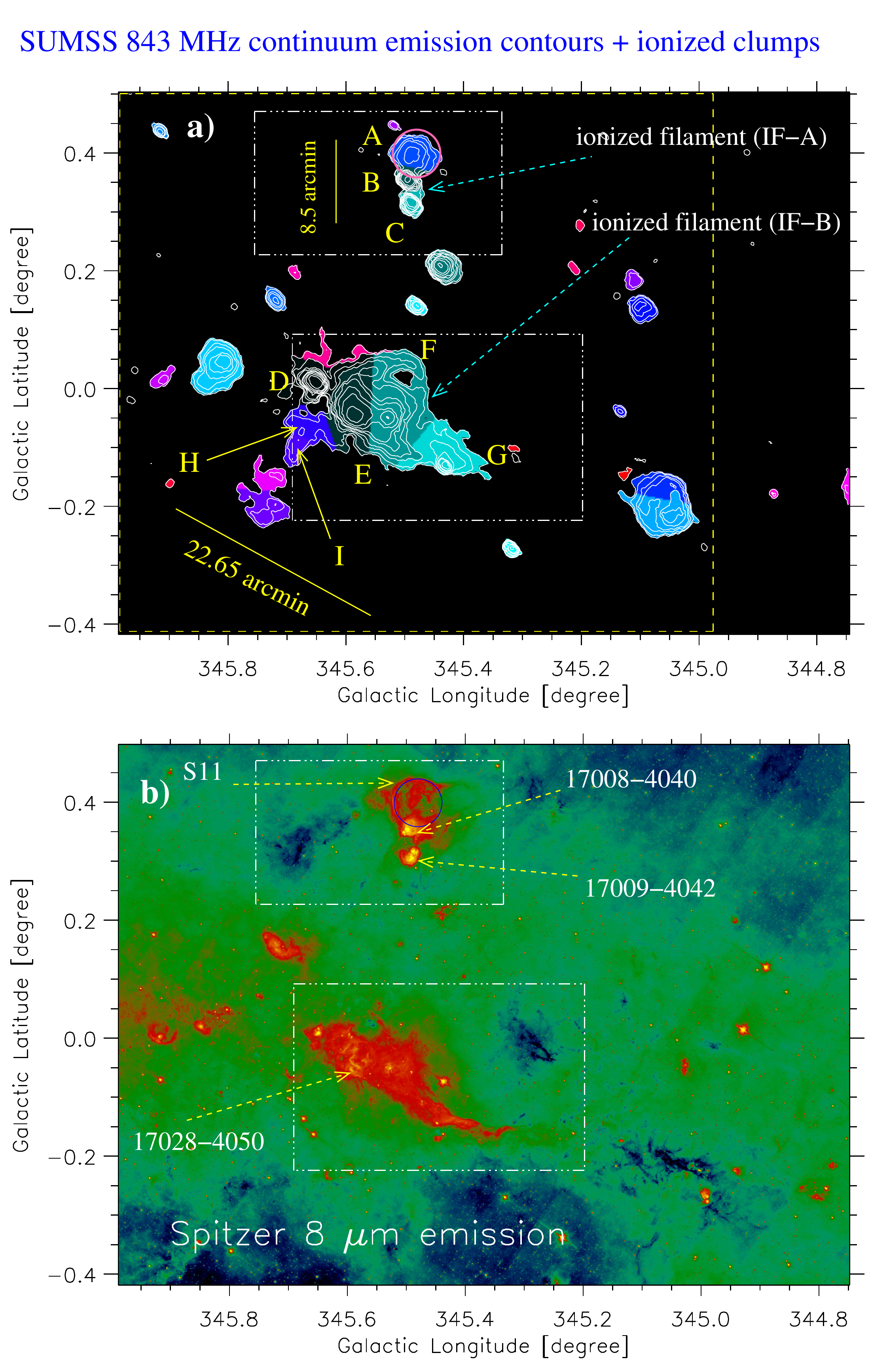}
\caption{a) The panel shows the SUMSS 843 MHz radio continuum contours toward an area of $\sim$74\rlap.{$'$}64 $\times$ 55\rlap.{$'$}02 (centered at {\it l} = 345$\degr$.3693; {\it b} = 0$\degr$.0391). The radio continuum contours are plotted with the levels of 13, 20, 40, 60, 100, 130, 200, 300, 
and 550 mJy beam$^{-1}$. The background map displays a clumpfind decomposition of the 843 MHz continuum emission, 
highlighting the spatial boundary of ionized clumps. A dashed box highlights an area presented in Figures~\ref{fig2}a--\ref{fig2}d. 
b) The panel presents the {\it Spitzer} image at 8.0 $\mu$m. 
A big circle shows the location of the MIR bubble S11 (average radius $\sim$2\rlap.{$'$}43) in both panels. 
In each panel, two dotted-dashed boxes highlight areas hosting elongated morphologies.}
\label{fig1}
\end{figure}

The data sets utilized in this work are listed in Table~\ref{tab1}, and were obtained 
toward our selected area around {\it l} = 345$\degr$.5 as presented in Figure~\ref{fig1}a. 
In this paper, we used the Gaia early data release 3 \citep[EDR3;][]{gaia21,fabricius21} based photogeometric distances (``rpgeo'') of point-like sources from \citet{bailer21}.  
Based on the analysis of the {\it Herschel} continuum images at 70--500 $\mu$m \citep{Molinari10a}, the {\it Herschel} temperature and column density maps (resolution $\sim$12$''$) were constructed for the {\it EU-funded ViaLactea project} \citep{Molinari10b}. The Bayesian {\it PPMAP} procedure \citep{marsh15,marsh17} was applied to obtain these {\it Herschel} maps. 
The {\it Herschel} temperature map is used in this paper. 

The SEDIGISM $^{13}$CO/C$^{18}$O(J = 2--1) line data \citep[beam size $\sim$30$''$; pixel-scale $\sim$9\rlap.{$''$}5; rms $\sim$0.8--1.0~K;][]{schuller17,schuller21} and 
the Mopra $^{13}$CO(J = 1--0) line data \citep[beam size $\sim$36$''$; pixel-scale $\sim$30$''$; rms $\sim$0.5~K;][]{braiding18} are examined in this paper. 
These line data were smoothed with a Gaussian function having a width of 3 pixels. 
The smoothing process gives the resultant angular resolutions of the SEDIGISM $^{13}$CO/C$^{18}$O(J = 2--1) line data and 
Mopra $^{13}$CO(J = 1--0) line data to be $\sim$41\rlap.{$''$}4 and $\sim$96\rlap.{$''$}9, respectively. The Mopra $^{13}$CO(J = 1--0) line data are not available for our entire selected target area, but these observations (i.e., $\sim$60\rlap.{$'$}6 $\times$ 54\rlap.{$'$}9; centered at {\it l} = 345$\degr$.4801; {\it b} = 0$\degr$.0442) cover both the ionized filaments. 
Apart from the $^{13}$CO/C$^{18}$O line data, we also studied the N$_{2}$H$^{+}$(1--0) line data from 
the MALT90 survey \citep[beam size $\sim$38$''$; rms $\sim$0.2~K;][]{foster11,jackson13} 
mainly toward i17008 and i17009. 
\begin{table*}
% \tiny
%\scriptsize
\small
\setlength{\tabcolsep}{0.1in}
\centering
\caption{List of observational surveys utilized in this work.}
\label{tab1}
\begin{tabular}{lcccr}
\hline 
  Survey  &  Wavelength/Frequency/line(s)       &  Resolution ($\arcsec$)        &  Reference \\   
\hline
\hline 
SUMSS                 &843 MHz                     & $\sim$45        &\citet{bock99}\\
SEDIGISM&  $^{13}$CO/C$^{18}$O (J = 2--1) & $\sim$30        &\citet{schuller17}\\
Mopra Galactic Plane CO survey&  $^{12}$CO/$^{13}$CO/C$^{18}$O (J = 1--0) &   $\sim$36        &\citet{braiding18}\\
Millimeter Astronomy Legacy Team Survey at 90 GHz (MALT90)                 & molecular lines near 90 GHz                     & $\sim$38        &\citet{jackson13}\\
ATLASGAL                 &870 $\mu$m                     & $\sim$19.2        &\citet{schuller09}\\
{\it Herschel} Infrared Galactic Plane Survey (Hi-GAL)                              &70--500 $\mu$m                     & $\sim$5.8--37         &\citet{Molinari10a}\\
{\it Spitzer} MIPS Inner Galactic Plane Survey (MIPSGAL)                                         &24 $\mu$m                     & $\sim$6         &\citet{carey05}\\ 
Wide Field Infrared Survey Explorer (WISE) & 12 $\mu$m                   & $\sim$6           &\citet{wright10}\\ 
{\it Spitzer}-GLIMPSE       &3.6--8.0  $\mu$m                   & $\sim$2           &\citet{benjamin03}\\
\hline          
\end{tabular}
\end{table*}
\section{Results}
\label{sec:data}
\subsection{Physical environments around {\it l} = 345$\degr$.5} 
\label{sec:morphx}
This section focuses to probe the distribution of dust emission, clumps, ionized emission, molecular gas, and embedded protostars around {\it l} = 345$\degr$.5, which allows us to identify various emission structures, their physical association, and signatures of star formation. Such an investigation is very useful to probe the physical conditions around {\it l} = 345$\degr$.5. 
\subsubsection{Ionized clumps toward elongated ionized filaments}
\label{sec:morph}
In order to explore the ionized clumps/H\,{\sc ii} regions and the ionized filaments, we have 
employed the radio 843 MHz continuum map in the direction of our selected area around {\it l} = 345$\degr$.5. 
As mentioned earlier, the spatial appearance of the 843 MHz continuum emission enabled us to trace two elongated ionized filaments around {\it l} = 345$\degr$.5 (see Figure~\ref{fig1}a), which are designated as IF-A (extent $\sim$8\rlap.{$'$}5) and IF-B (extent $\sim$22\rlap.{$'$}65). In Figure~\ref{fig1}b, the extended emission traced in the 8.0 $\mu$m image can be also observed toward both the ionized filaments. The locations of at least three IRAS sources (i.e., i17008, i17009, and IRAS 17028-4050) and a previously known mid-infrared (MIR) bubble S11 \citep[l = 345$\degr$.48; b = 0$\degr$.399;][]{churchwell06} are indicated in Figure~\ref{fig1}b. 
The ionized filament IF-A, hosting i17008, i17009, and the bubble S11 is traced in the northern direction, while the filament IF-B is depicted in the southern direction. 

In Figures~\ref{fig1}a and~\ref{fig1}b, the location of the bubble S11 is also marked by a circle (average radius = 2\rlap.{$'$}43). 
This bubble \citep[distance $\sim$2.0 kpc;][]{watson10} was identified as a complete/closed ring or a probable enclosed central star cluster with an average radius and thickness of 2\rlap.{$'$}43 and 0\rlap.{$'$}39, respectively \citep[see][]{churchwell06,hanaoka20}. From the previous work of \citet{dewangan18}, we find one ATLASGAL dust continuum clump at 870 $\mu$m \citep[i.e., clm3; M$_\mathrm{clump}$ $\sim$600 M$_{\odot}$; $T_\mathrm{d}$ $\sim$16.5 K; V$_\mathrm{lsr}$ = $-$15.8 km s$^{-1}$; d $\sim$2.4 kpc; see also][]{urquhart18} toward the bubble S11. Based on the previously reported V$_\mathrm{lsr}$ values toward the dust clumps associated with i17008 and i17009 (i.e., clm1 (V$_\mathrm{lsr}$ = $-$17 km s$^{-1}$) and clm2 (V$_\mathrm{lsr}$ = $-$17.3 km s$^{-1}$)), we may suggest that the clump clm3 (V$_\mathrm{lsr}$ = $-$15.8 km s$^{-1}$) appears to be redshifted compared to the other two clumps. But, it requires further investigation using molecular line data. 

In addition to the elongated ionized structures, several peaks are visually seen in the radio continuum map. 
Hence, we employed the {\it clumpfind} IDL program \citep{williams94} to identify the ionized clumps/H\,{\sc ii} regions from the SUMSS 843 MHz continuum map. The {\it clumpfind} program also allows us to obtain the total flux density ($S_\mathrm{\nu}$) of each selected ionized clump/H\,{\sc ii} region. 
However, we have labeled nine ionized clumps (i.e., A--I), which are distributed mainly toward IF-A and IF-B (see Figure~\ref{fig1}a).
In the direction of IF-A, the ionized clumps A, B, and C are found toward the bubble S11, i17008, and i17009, respectively. 
Six ionized clumps (i.e., D--I) are labeled toward IF-B. 
In general, the observed flux value is used to compute the number of 
Lyman continuum photons $N_\mathrm{UV}$ of an ionized clump/H\,{\sc ii} region, 
and in this relation, one can use the following equation \citep{matsakis76}:
\begin{equation}
\begin{split}
N_\mathrm{UV} (s^{-1}) = 7.5\, \times\, 10^{46}\, \left(\frac{S_\mathrm{\nu}}{\mathrm{Jy}}\right)\left(\frac{D}{\mathrm{kpc}}\right)^{2} 
\left(\frac{T_\mathrm{e}}{10^{4}\mathrm{K}}\right)^{-0.45} \\ \times\,\left(\frac{\nu}{\mathrm{GHz}}\right)^{0.1}
\end{split}
\end{equation}
\noindent where $S_\mathrm{\nu}$ (in Jy) is the total flux 
density of the H\,{\sc ii} region, $D$ is the distance in kpc, $T_\mathrm{e}$ is the electron temperature, and $\nu$ is the frequency in GHz. 
With the help of the equation~1, $T_\mathrm{e}$ = 10$^{4}$~K, and $D$ = [1.4 kpc, 2.4 kpc; see Section~\ref{sec:morph2} for more details], we determine $\log{N_\mathrm{UV}}$ of each ionized clump marked in Figure~\ref{fig1}a. 
Using the reference of \citet{panagia73}, these clumps (i.e., A--I) are found to be powered by massive B0.5V-O9.5V type stars. 
Furthermore, following the equations and analysis adopted in \citet{dewangan17a}, the typical value of the initial particle number density of the ambient neutral gas ($n_\mathrm{0}$ = 10$^{3}$ (10$^{4}$) cm$^{-3}$) leads a range of dynamical age of the ionized clumps (i.e., A--I) to be $\sim$0.1--0.3 (0.3--1) Myr. The analysis shows the presence of massive stars in both the ionized filaments, which are distributed along the filaments.
\subsubsection{Distribution of dust clumps, protostars, and molecular gas toward ionized filaments}
\label{sec:morph2}
In this section, we explore the multi-wavelength data sets to examine the embedded dust/molecular structures and protostars/young stellar objects (YSOs) against the ionized features around {\it l} = 345$\degr$.5.

\begin{figure*}
\center
\includegraphics[width=17cm]{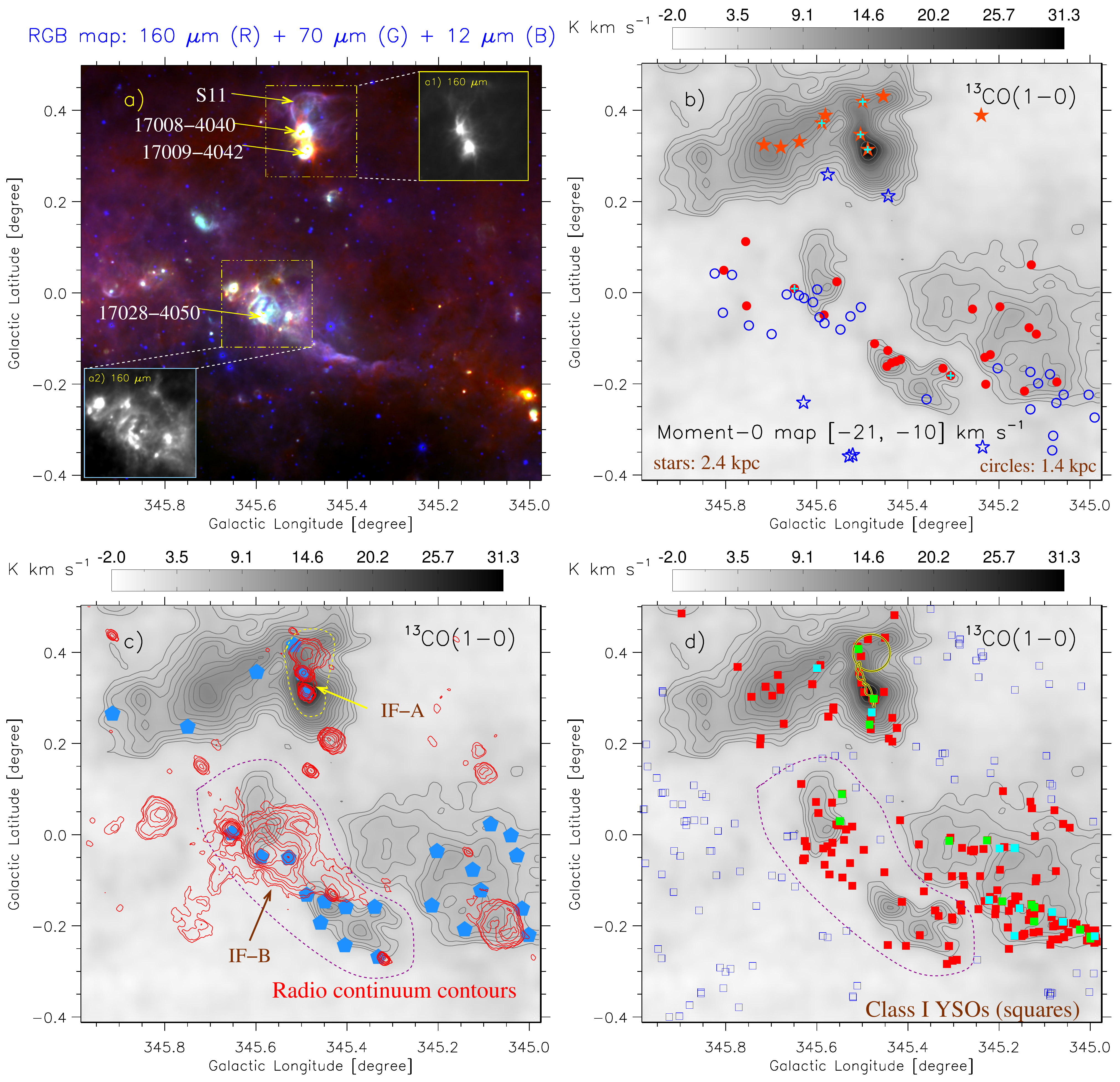}
\caption{a) The panel shows a 3-color composite map made using the {\it Herschel} 160 $\mu$m (in red), {\it Herschel} 70 $\mu$m (in green), and WISE 12 $\mu$m (in blue) images. 
Using the {\it Herschel} 160 $\mu$m image, a zoomed-in view around i17008 is shown in the inset on the top right, while a zoomed-in view around IRAS 17028-4050 is presented in the inset on the bottom left. 
b) The panel presents the Mopra $^{13}$CO(J = 1--0) integrated intensity emission map 
at [$-$21, $-$10] km s$^{-1}$. The Mopra $^{13}$CO emission contours are also plotted with the 
levels of 4.3, 5, 6, 7, 8, 9, 10, 11, 12, 13, 16, 20, and 25 K km s$^{-1}$. 
The Mopra $^{13}$CO line intensity is presented in terms of the antenna temperature. The molecular map is also overlaid with the positions of ATLASGAL clumps at 870 $\mu$m \citep[see stars and circles; from][]{urquhart18}. All circles show the ATLASGAL clumps located at d = 1.4 kpc, while stars represent 
the ATLASGAL clumps situated at d = 2.4 kpc. All the filled symbols show the ATLASGAL clumps 
with V$_\mathrm{lsr}$ of [$-$24, $-$8] km s$^{-1}$, while all the open symbols highlight the ATLASGAL clumps with V$_\mathrm{lsr}$ of [$-$7, 0] km s$^{-1}$ and [$-$30, $-$25] km s$^{-1}$. Plus symbols show the ATLASGAL clumps associated with the $^{13}$CO outflows \citep[see][for more details]{yang22}.  
c) Overlay of the SUMSS 843 MHz radio continuum contours (in red; see Figure~\ref{fig1}a) on the Mopra $^{13}$CO intensity emission map. 
Filled pentagons represent the positions of 26 IRAS sources located in the direction of elongated morphologies. 
Arrows indicate two ionized filaments (IF-A and IF-B) in the panel. d) Overlay of Class~I YSOs (see filled and open squares) on the Mopra $^{13}$CO intensity emission map. 
This paper mainly focuses on the Class~I YSOs, which are indicated by filled squares. Gaia optical counterparts of some selected Class~I YSOs are available, and are highlighted by cyan filled squares (d = [1.6, 1.96] kpc) and green filled squares (d = [2.0, 2.6] kpc).     
An elongated filament and the bubble S11 are highlighted by a solid curve and a big circle, respectively. 
The elongated filament hosts the sites i17008 and i17009, and was traced using the 870 $\mu$m continuum map \citep[see][]{dewangan18}. 
In panels ``c" and ``d", the Mopra molecular emission contours are also shown as presented in Figure~\ref{fig2}b, 
and the extent of the elongated cloud hosting IF-B is indicated by a dotted curve (in magenta).}  
\label{fig2}
\end{figure*}

Figure~\ref{fig2}a shows a 3-color composite map made using the {\it Herschel} 160 $\mu$m (in red), {\it Herschel} 70 $\mu$m (in green), and WISE 12 $\mu$m (in blue) images. 
Filamentary structures and bubble-like features are clearly visible in the infrared images, which trace the dust emission.  
The inset on the top right presents an area -- hosting i17008, i17009, 
and S11-- in the zoomed-in view using the {\it Herschel} 160 $\mu$m image, showing the earlier reported one HFS toward i17008 and i17009. The inset on the bottom left displays an area hosting IRAS 17028-4050 
in the zoomed-in view using the {\it Herschel} 160 $\mu$m image, showing the infrared bubble-like features.

We have examined the Mopra $^{13}$CO(J = 1--0) emission in a velocity range of [$-$21, $-$10] km s$^{-1}$ to study the distribution of molecular gas. Figure~\ref{fig2}b displays the Mopra $^{13}$CO(J = 1--0) integrated emission map (moment-0) overlaid with the positions of the ATLASGAL clumps at 870 $\mu$m (see circles and stars). 
Note that \citet{urquhart18} also determined the reliable velocities and distances of the ATLASGAL clumps, which can be used to study the physical connection of different sub-regions in a given large area.  
In Figure~\ref{fig2}b, the ATLASGAL clumps marked by stars and circles are located at a distance of 2.4 kpc and 1.4 kpc, respectively \citep[see][for more details]{urquhart18}. The ATLASGAL clumps associated with the $^{13}$CO outflows are highlighted by plus symbols (in cyan; see Figure~\ref{fig2}b). 
This information is taken from \citet{yang22}, who listed the detection of the $^{13}$CO outflows. They also provided the velocity ranges of the $^{13}$CO(J = 2--1) red and blue wing-like velocity components toward the ATLASGAL clumps associated with outflows using the SEDIGISM $^{13}$CO(J = 2--1) and C$^{18}$O(J = 2--1) line data.   
All filled symbols show the clumps with V$_\mathrm{lsr}$ of [$-$24, $-$8] km s$^{-1}$. 
On the other hand, open circles and stars represent clumps with V$_\mathrm{lsr}$ of [$-$7, 0] km s$^{-1}$ and [$-$30, $-$25] km s$^{-1}$, which may not be associated with IF-A and IF-B. 
Note that in the direction of both the ionized filaments IF-A and IF-B, the Mopra molecular gas is depicted in the same velocity range of 
[$-$21, $-$10] km s$^{-1}$. On the basis of the distances to the ATLASGAL clumps, IF-A and IF-B appear to be located at a distance of 2.4 kpc and 1.4 kpc, respectively. 
\begin{figure}
\center
\includegraphics[width=8.5cm]{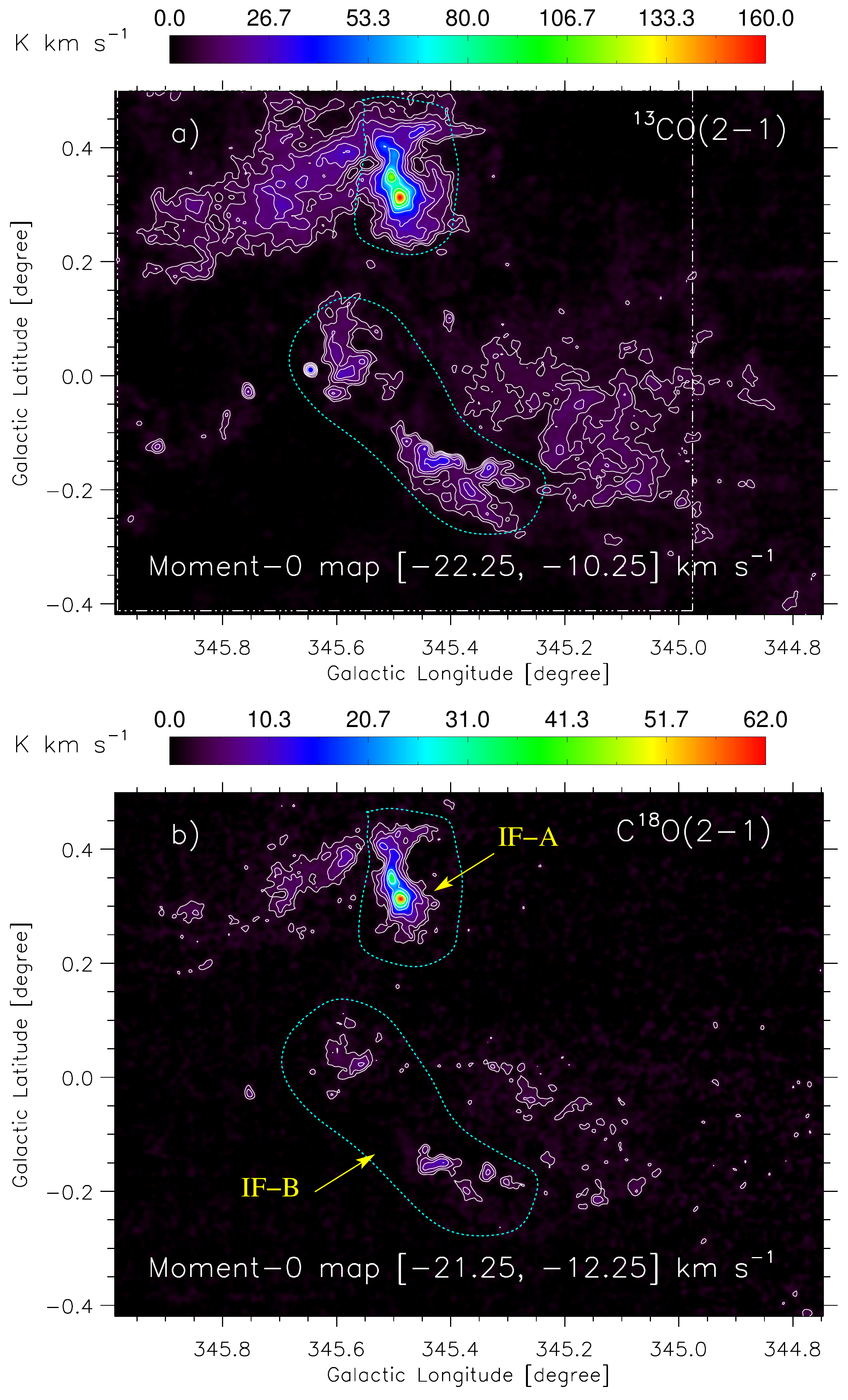}
\caption{a) SEDIGISM $^{13}$CO(J = 2--1) map of intensity (moment-0) in the direction of our selected target area (see Figure~\ref{fig1}a). 
The molecular emission is integrated from $-$22.25 to $-$10.25 km s$^{-1}$. 
The $^{13}$CO emission contours are also shown with the levels of 7, 10, 15, 20, 28, 48, 70, 90, and 120 K km s$^{-1}$. 
A dotted-bashed box highlights an area covered by Mopra line data (see Figure~\ref{fig2}b). 
b) SEDIGISM C$^{18}$O(J = 2--1) moment-0 map. The molecular emission is integrated from $-$21.25 to $-$12.25 km s$^{-1}$. 
The C$^{18}$O emission contours are also shown with the levels of 2.6, 4.3, 7, 11, 20, 28, and 42 K km s$^{-1}$. In each panel, dotted curves indicate the elongated parent clouds of IF-A and IF-B.} 
\label{fig3}
\end{figure}

In Figure~\ref{fig2}c, the distribution of ionized emission (red contours) and the positions of 26 IRAS sources (filled pentagons) are presented against the molecular emission.
The distribution of molecular gas traces a continuous structure toward IF-A,
but a deficiency of molecular gas is seen toward the central part of IF-B (see Figure~\ref{fig2}c).
At least one molecular condensation is found toward both the ends of IF-B.
From Figure~\ref{fig2}a, we infer a continuous structure in the infrared images toward IF-B.
Based on multi-wavelength images and distribution of the ATLASGAL clumps, we suggest that there was an elongated molecular filament (see a dotted curve in Figure~\ref{fig2}c), which has been eroded by the impact of massive stars located at the center of IF-B.

We examined the {\it Spitzer}-GLIMPSE photometric data at 3.6--5.8 $\mu$m, which allowed us to identify younger protostars (i.e., Class~I YSOs) in our selected target area. The photometric magnitudes of point-like sources at {\it Spitzer} 3.6--5.8 $\mu$m bands were collected from the GLIMPSE-I Spring' 07 highly reliable catalog \citep{benjamin03}. Class~I YSOs are selected using the infrared color 
conditions (i.e., [4.5]$-$[5.8] $\ge$ 0.7 mag and [3.6]$-$[4.5] $\ge$ 0.7 mag) described in \citet{hartmann05} and \citet{getman07}. 
In Figure~\ref{fig2}d, we display the positions of Class~I YSOs overlaid on the Mopra $^{13}$CO map. An elongated filament traced in the 870 $\mu$m continuum map \citep[see][]{dewangan18} and 
the location of the bubble S11 are also indicated by a curve and a big circle (average radius = 2\rlap.{$'$}43 or 1.7 pc at a distance of 2.4 kpc), respectively. 

In this work, we focus on only those Class~I YSOs, which are distributed toward the clumpy structures in the clouds (see filled squares in Figure~\ref{fig2}d). Such selection is considered by the visual inspection of the molecular gas and dust emissions (see the {\it Herschel} 160 $\mu$m emission in Figure~\ref{fig2}a and the ATLASGAL clumps in Figure~\ref{fig2}b). Several Class~I YSOs appear to be located outside the molecular cloud boundary, which is traced using the Mopra $^{13}$CO emission contour with a level of 4.3 K km s$^{-1}$ (see Figure~\ref{fig2}d). Hence, such Class~I YSOs are unlikely to be part of the target clouds (see open squares in Figure~\ref{fig2}d). 
Therefore, we have not made any attempt to study the Class~I YSOs, which are highlighted by open squares in Figure~\ref{fig2}d.  
In order to get distance information of the selected protostars (see filled squares in Figure~\ref{fig2}d), we examined point-like sources in the Gaia EDR3 catalog \citep{gaia21,bailer21}.
 
In the direction of the clouds traced in Figure~\ref{fig2}c, the distance distribution of Gaia point-like sources peaks around a distance of 2.5 kpc (not shown here). It is expected that the optical counterparts of the selected Class~I YSOs may be faint and/or may not be detected in the Gaia EDR3 catalog. We find optical counterparts of some Class~I YSOs toward the clouds.  
The distances of some of these sources are in agreement with the dust clumps. In this relation, we have displayed the GAIA optical counterparts of Class~I YSOs by cyan filled squares (d = [1.6, 1.96] kpc) 
and green filled squares (d = [2.0, 2.6] kpc) in Figure~\ref{fig2}d. 
This particular analysis favors that the Class~I YSOs spatially seen toward the clumpy structures in the clouds are likely to be physically connected with the ionized emission, dust clumps and molecular material. Hence in other words, in the direction of IF-A, we find an obvious correspondence among the ionized emission, dust clumps and molecular material, where Class~I YSOs are distributed.  
From Figures~\ref{fig2}b and~\ref{fig2}d, dust clumps and Class~I YSOs are also seen toward the central part of IF-B, 
where a deficiency of molecular gas is found.
 
In general, an average age of Class~I YSOs is reported to be $\sim$0.44 Myr \citep{evans09}. 
Overall, the early phases of star formation activities and the presence of massive stars are evident toward the parent clouds of the ionized filaments (see Figure~\ref{fig2}d).
\begin{figure}
\center
\includegraphics[width=8.5cm]{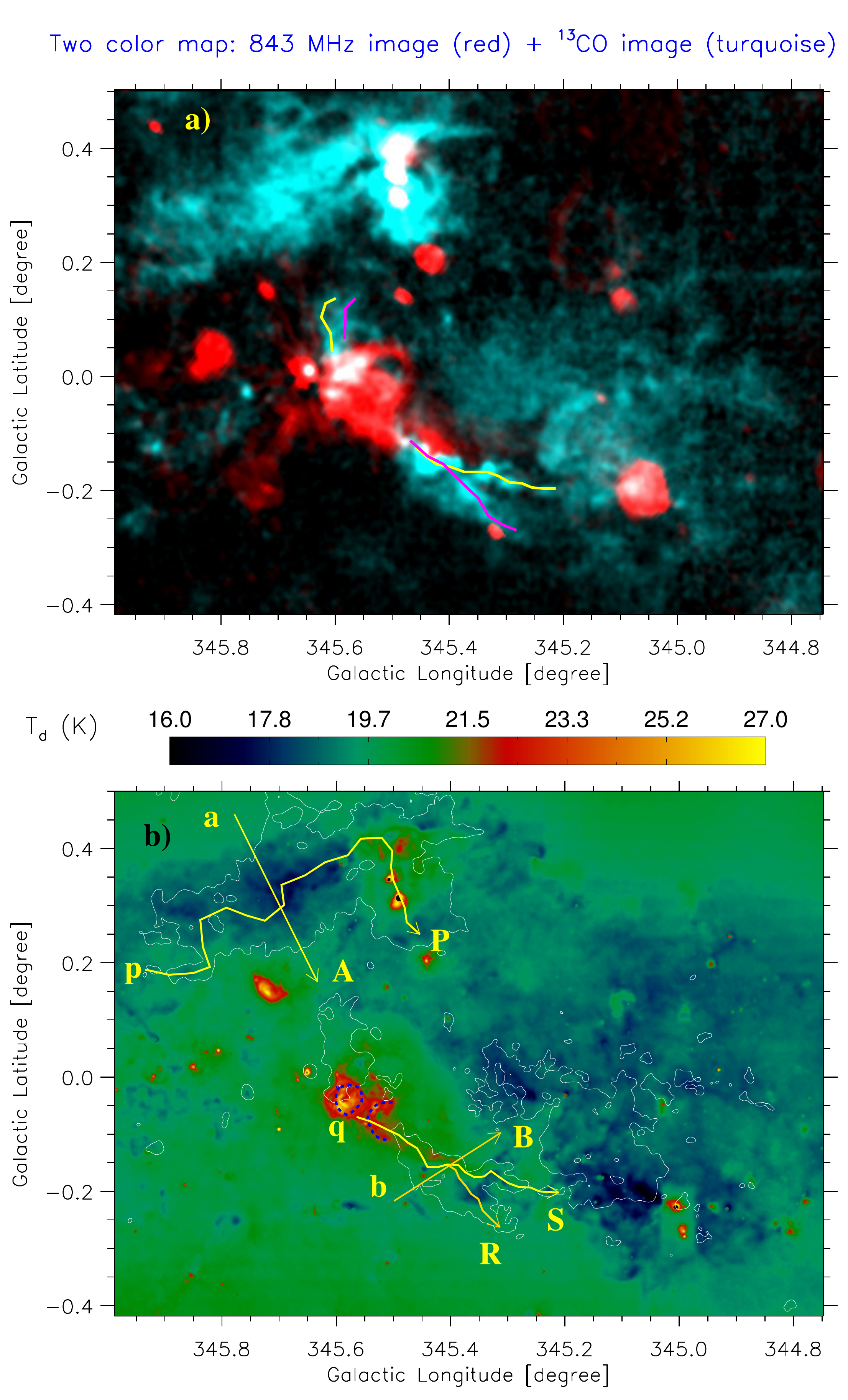}
\caption{a) The panel shows a two-color composite image made using the SUMSS 843 MHz continuum emission (in red) 
and the $^{13}$CO(J = 2--1) moment-0 map (in turquoise color; see Figure~\ref{fig3}a). 
In the direction of IF-B, at least two elongated molecular filaments are indicated by curves. 
b) The panel displays the overlay of the $^{13}$CO(J = 2--1) emision contour (see Figure~\ref{fig3}a) 
on the {\it Herschel} temperature map. Two arrows (i.e., aA and bB) and three curves (i.e., pP, qR and qS) are 
indicated in the panel, where the position-velocity diagrams are produced (see Figures~\ref{fig5}e,~\ref{fig5}f, and~\ref{fig6}). 
Blue dashed curves highlight bubble-like structures in the temperature map.}
\label{fig4}
\end{figure}
\subsection{SEDIGISM $^{13}$CO(J = 2--1) and C$^{18}$O(J = 2--1) emission}
\label{sec:gasmorphb}
In this section, we study the kinematics of molecular gas around {\it l} = 345$\degr$.5, allowing us to examine gas velocity structures. Such knowledge is essential to probe the ongoing physical processes toward the selected target area.
\subsubsection{Molecular clouds hosting IF-A and IF-B}
\label{sec:gasmorphbx}
Here we study the spatial and velocity distribution of the SEDIGISM $^{13}$CO(J = 2--1) and C$^{18}$O(J = 2--1) emission in the area shown in Figure~\ref{fig1}a. Figures~\ref{fig3}a and~\ref{fig3}b present the $^{13}$CO(J = 2--1) and C$^{18}$O(J = 2--1) integrated maps and contours, respectively, where enclosed regions indicate the areas around the ionized filaments. 
The integrated intensity or moment-0 maps of $^{13}$CO(J = 2--1) and C$^{18}$O(J = 2--1) are produced using velocity intervals of [$-$22.25, $-$10.25] and [$-$21.25, $-$12.25] km s$^{-1}$, respectively. Similar morphologies of clouds are evident in the Mopra $^{13}$CO(J = 1--0) and the SEDIGISM $^{13}$CO(J = 2--1) maps. However, one can keep in mind that the SEDIGISM molecular maps (beam size $\sim$41\rlap.{$''$}4) provide more insight into the clouds due to its relatively better resolution compared to the Mopra line 
data (beam size $\sim$96\rlap.{$''$}9). In Figure~\ref{fig3}b, the C$^{18}$O(J = 2--1) emission enables us to depict denser parts in the molecular clouds 
as traced by the $^{13}$CO(J = 2--1) emission (see areas of enclosed regions in Figures~\ref{fig3}a and~\ref{fig3}b).  

In Figure~\ref{fig4}a, we display a two-color composite map made using the SUMSS 843 MHz continuum map (in red) and 
the SEDIGISM $^{13}$CO(J = 2--1) map (in turquoise). 
IF-A is embedded in the filamentary molecular cloud, which is distinctly traced in 
the C$^{18}$O(J = 2--1) map (see Figure~\ref{fig3}b). 
The color composite map indicates the destruction of the central part of an elongated molecular (i.e., $^{13}$CO(J = 2--1) and C$^{18}$O(J = 2--1)) structure, 
where IF-B is spatially traced. The color composite map also hints at the presence of two filamentary molecular clouds toward IF-B, which are indicated by two curves in Figure~\ref{fig4}a. Figure~\ref{fig4}b displays the {\it Herschel} temperature map overlaid with 
the SEDIGISM $^{13}$CO(J = 2--1) emission contour. 
The areas around i17008 and i17009 are saturated in the {\it Herschel} temperature map. 
Warm dust emission ($T_\mathrm{d}$ $>$ 21~K) is evident toward both the ionized filaments. 
In the direction of the ionized clumps ``E" and ``F" in IF-B (see Figure~\ref{fig1}a), at least 
two bubble-like structures are observed in the {\it Herschel} temperature map (see blue dashed curves in Figure~\ref{fig4}b), 
where the molecular gas depression is found.

\begin{figure*}
\center
\includegraphics[width=13cm]{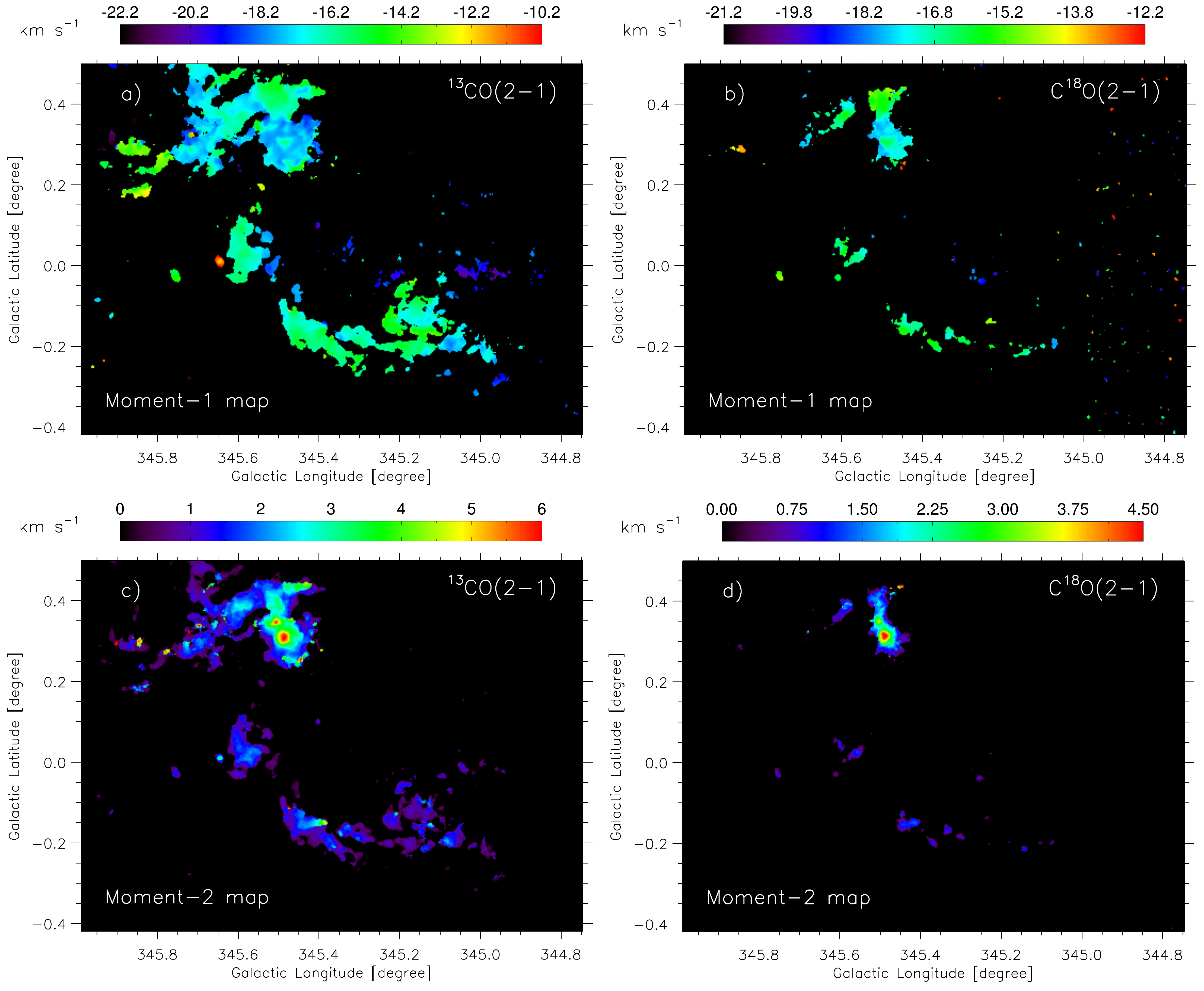}
\includegraphics[width=14cm]{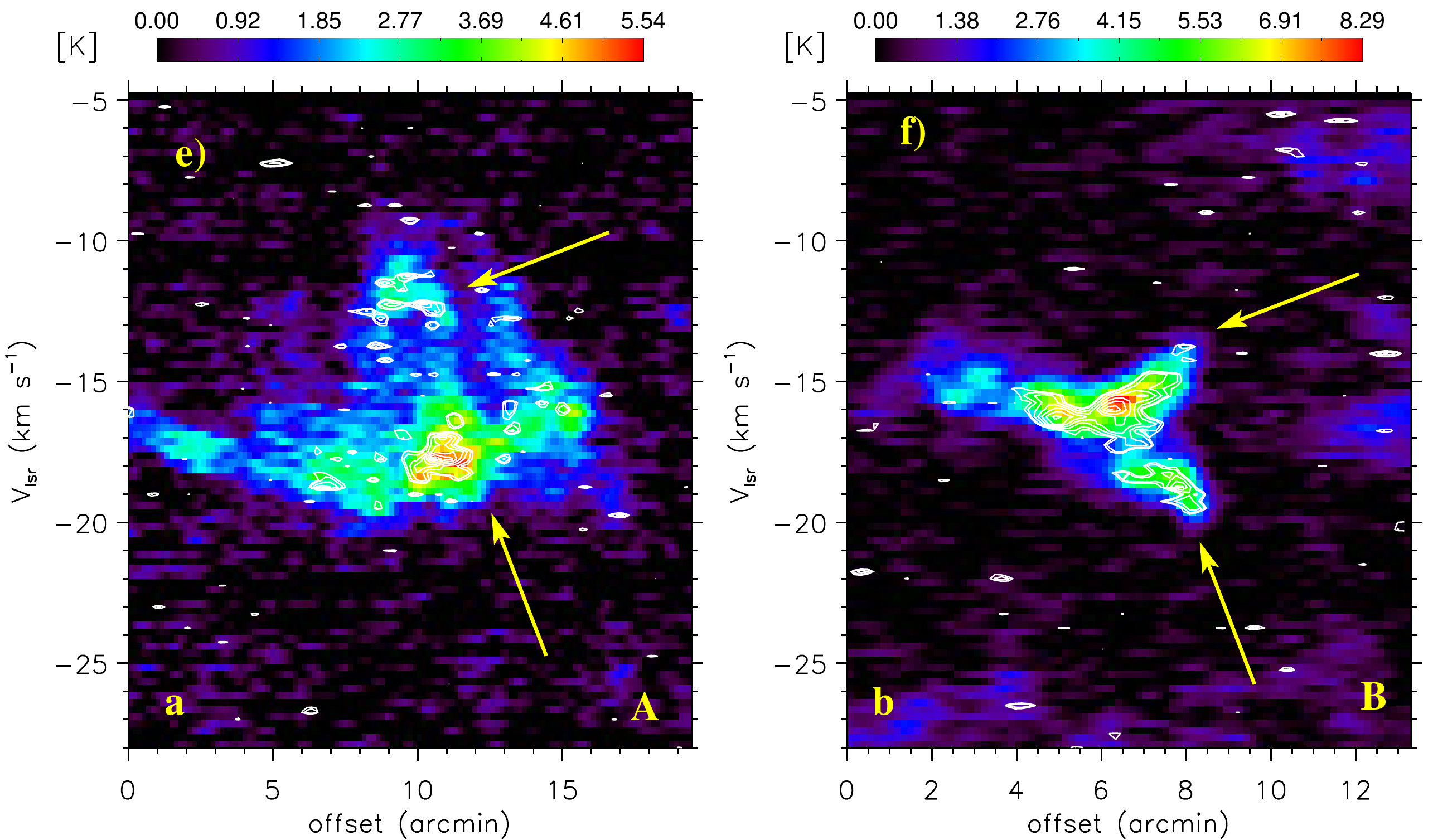}
\caption{a) $^{13}$CO(J = 2--1) moment-1 map of our selected target area (see a dotted-dashed box in Figure~\ref{fig1}a). 
b) C$^{18}$O(J = 2--1) moment-1 map. 
c) $^{13}$CO(J = 2--1) moment-2 map. 
d) C$^{18}$O(J = 2--1) moment-2 map. 
e) Position-velocity diagram of $^{13}$CO(J = 2--1) along the arrow ``aA" (see Figure~\ref{fig4}b). 
White contours are the C$^{18}$O(J = 2--1) emission. 
The levels of the C$^{18}$O(J = 2--1) emission contours are (0.35, 0.4, 0.5, 0.6, 0.7, 0.8, 0.9, 0.98) $\times$ 2.37 K. 
f) Position-velocity diagram of $^{13}$CO(J = 2--1) along the arrow ``bB" (see Figure~\ref{fig4}b). 
White contours are the C$^{18}$O(J = 2--1) emission. 
The levels of the C$^{18}$O(J = 2--1) emission contours are (0.35, 0.4, 0.5, 0.6, 0.7, 0.8, 0.9, 0.98) $\times$ 2.23 K. 
In panels ``e" and ``f", arrows indicate structures with different velocities.}
\label{fig5}
\end{figure*}

Figures~\ref{fig5}a and~\ref{fig5}b present the line velocity/velocity field/moment-1 map of the SEDIGISM $^{13}$CO(J = 2--1) and C$^{18}$O(J = 2--1) emission, respectively. Both these moment-1 maps indicate a noticeable velocity spread toward the clouds 
associated with both the ionized filaments, where higher values of line widths ($>$ 1.5 km s$^{-1}$) are found in 
the intensity-weighted line width maps (moment-2) of $^{13}$CO(J = 2--1) and C$^{18}$O(J = 2--1) (see Figures~\ref{fig5}c and~\ref{fig5}d). 

\begin{figure}
\center
\includegraphics[width=8.5 cm]{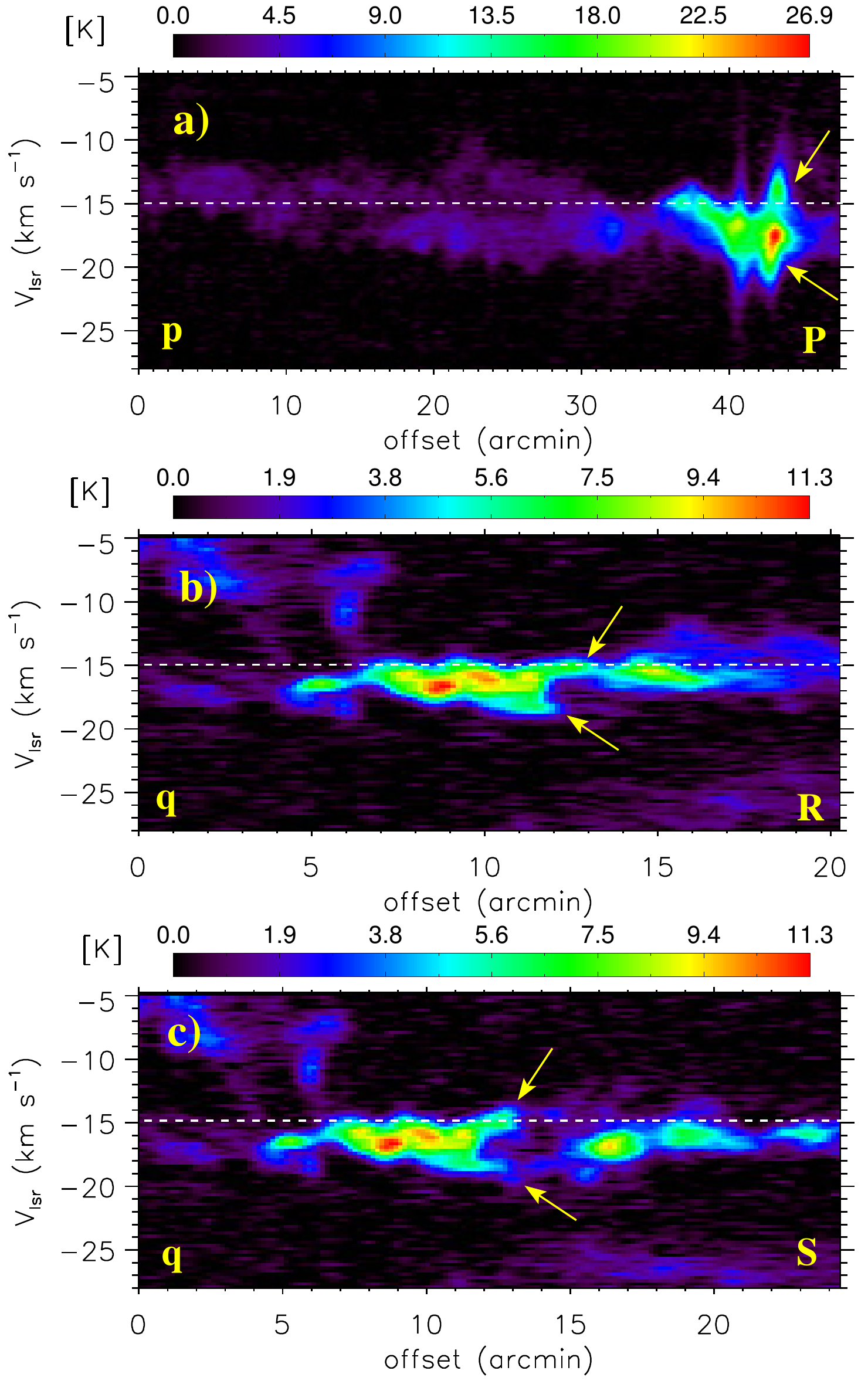}
\caption{Position-velocity diagrams of $^{13}$CO(J = 2--1) along the curves a) ``pP"; b) ``qR"; c) ``qS" (see Figure~\ref{fig4}b). 
Arrows indicate structures with different velocities.}
\label{fig6}
\end{figure}

Position-velocity diagrams of the $^{13}$CO(J = 2--1) emission along arrows ``aA" and ``bB" (see Figure~\ref{fig4}b) are presented in Figures~\ref{fig5}e and~\ref{fig5}f, respectively. 
The contours of the C$^{18}$O(J = 2--1) emission are also drawn in both the position-velocity diagrams. 
Along the direction of the arrow ``aA", the position-velocity diagram hints the existence of two velocity peaks/components around $-$11 and $-$17 km s$^{-1}$ (see arrows in Figure~\ref{fig5}e). 
Note that the arrow ``bB" passes through the overlapping areas of two filamentary molecular clouds in the direction of IF-B.
Along the arrow ``bB", we find at least two velocity peaks/components around $-$16 and $-$19 km s$^{-1}$, exhibiting the presence of two distinct filaments (see arrows in Figure~\ref{fig5}f). Figures~\ref{fig6}a,~\ref{fig6}b, and~\ref{fig6}c display 
position-velocity diagrams of the $^{13}$CO(J = 2--1) emission along curves ``pP", ``qR" and ``qS", which are marked in Figure~\ref{fig4}b. 
In the direction of i17008, an outflow \citep[blue wing: ($-$25.2, $-$19.8) km s$^{-1}$; red wing: ($-$14.8, $-$6.2) km s$^{-1}$;][]{yang22} is evident. We can also trace an outflow \citep[blue wing: ($-$22.8, $-$20.5) km s$^{-1}$; red wing: ($-$13.5, $-$9.8) km s$^{-1}$;][]{yang22} toward i17009. Apart from the outflow activity, in the direction of i17009, two velocity components (around $-$15 and $-$18 km s$^{-1}$) are also seen (see arrows in Figure~\ref{fig6}a).  
Furthermore, two distinct velocity peaks (around $-$15 and $-$18 km s$^{-1}$) are also evident along the curves ``qR" and ``qS" (see arrows in Figures~\ref{fig6}b and~\ref{fig6}c). 
\subsubsection{Zoomed-in view of molecular cloud associated with IF-A}
\label{sec:gasbx}
\begin{figure*}
\center
\includegraphics[width=\textwidth]{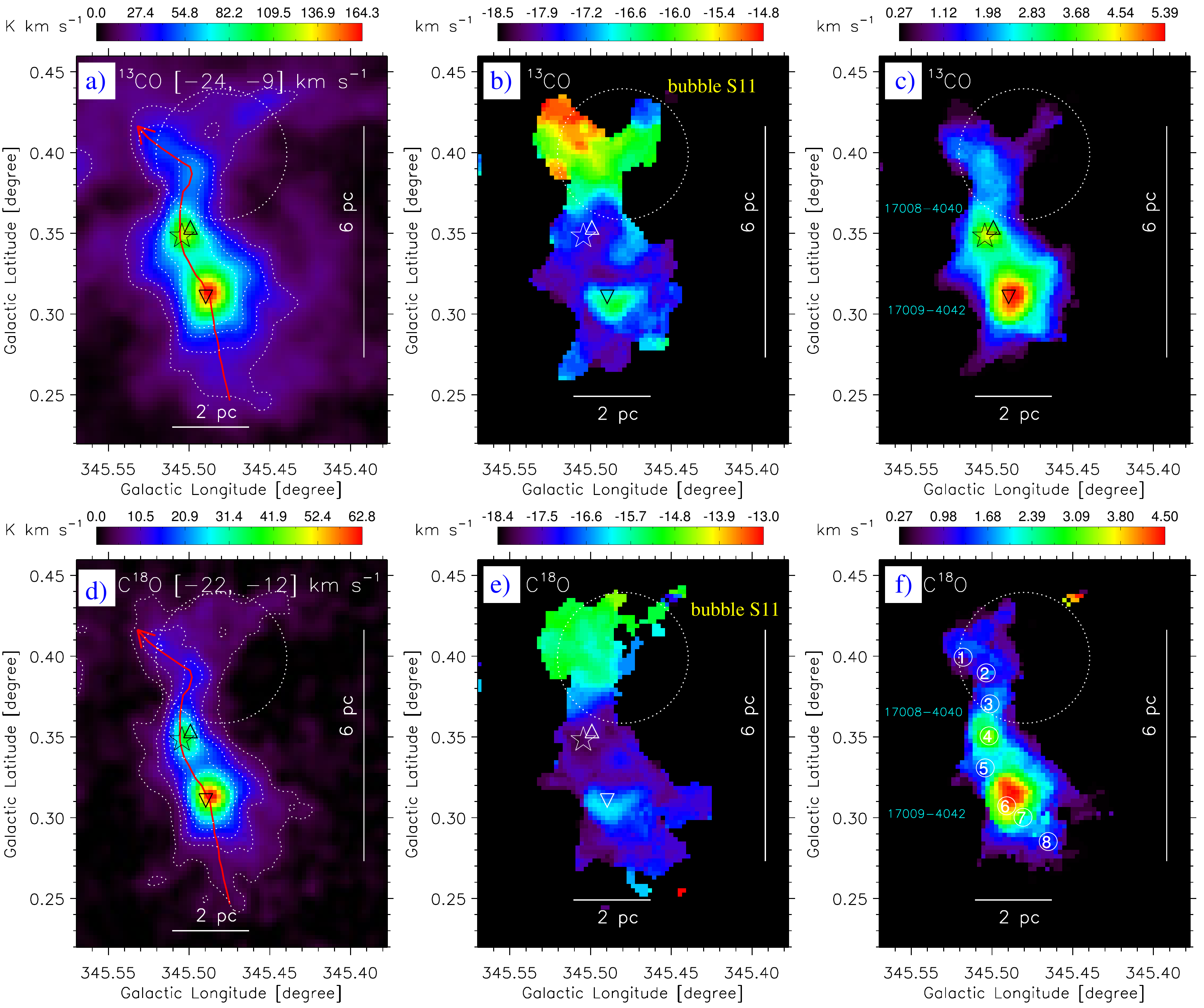}
\caption{A zoomed-in view of an area around i17008 using molecular line data sets. 
a) $^{13}$CO(J = 2--1) map of intensity (moment-0). 
The molecular emission is integrated from $-$24 to $-$9 km s$^{-1}$. 
The $^{13}$CO emission contours are also shown with the levels of (0.13, 0.2, 0.3, 0.4) $\times$ 164.3 K km s$^{-1}$. 
b) $^{13}$CO(J = 2--1) moment-1 map. c) $^{13}$CO(J = 2--1) moment-2 map.
d) C$^{18}$O(J = 2--1) moment-0 map. The molecular emission is integrated from $-$22 to $-$12 km s$^{-1}$. 
The C$^{18}$O(J = 2--1) emission contours are also shown with the levels of (0.06, 0.13, 0.2, 0.3, 0.4) $\times$ 62.8 K km s$^{-1}$. 
e) C$^{18}$O(J = 2--1) moment-1 map. f) C$^{18}$O(J = 2--1) moment-2 map. 
Eight circles are marked and labeled, where averaged spectra are produced in Figure~\ref{fig8}.
In panels ``a" and ``d", a curve (in red) is indicated, and position-velocity diagrams are extracted along it 
(see Figures~\ref{fig11}a and~\ref{fig11}b). In each panel, the positions of i17008, i17009, the MIR bubble S11, 
and 6.7 GHz MME are marked by triangle, upside down triangle, dotted circle, and star, respectively. 
In all panels, the scale bars derived at a distance of 2.4 kpc are shown.}
\label{fig7}
\end{figure*}

In the direction of an area hosting i17008, i17009, and S11 (or ionized clumps ``A--C"), Figures~\ref{fig7}a,~\ref{fig7}b, and~\ref{fig7}c display the moment-0, moment-1, and moment-2 maps of $^{13}$CO(J = 2--1), respectively. In Figures~\ref{fig7}d,~\ref{fig7}e, and~\ref{fig7}f, we show moment-0, moment-1, and moment-2 maps of C$^{18}$O(J = 2--1), respectively. 
The $^{13}$CO(J = 2--1) emission is integrated over a velocity range of [$-$24, $-$9] km s$^{-1}$ (see Figure~\ref{fig7}a), while the C$^{18}$O(J = 2--1) emission is integrated over a velocity range from $-$22 to $-$12 km s$^{-1}$ (see Figure~\ref{fig7}d). 
A dotted circle in each panel of Figure~\ref{fig7} highlights the location of the bubble S11. 
The moment-0 maps of $^{13}$CO(J = 2--1) and C$^{18}$O(J = 2--1) show the elongated filamentary structure (see Figures~\ref{fig7}a and~\ref{fig7}d), which has almost a similar morphology as seen in the ATLASGAL continuum map at 870 $\mu$m (see Figure~\ref{fig2}d).
However, the proposed HFSs are not very clearly seen in both the moment-0 maps. 

The moment-1 maps of $^{13}$CO(J = 2--1) and C$^{18}$O(J = 2--1) show a noticeable velocity difference/gradient toward the northern direction compared to the site i17009.  
In the direction of both the IRAS sites, we find higher values of line width ($>$ 2.5 km s$^{-1}$) (see Figures~\ref{fig7}c and~\ref{fig7}f). 
One can also find higher line width (i.e., 1--2.5 km s$^{-1}$) toward the bubble in the moment-2 maps 
of $^{13}$CO(J = 2--1) and C$^{18}$O(J = 2--1). 
The increased line widths can suggest star formation activities and/or the presence of multiple velocity components.

\begin{figure*}
\center
\includegraphics[width=\textwidth]{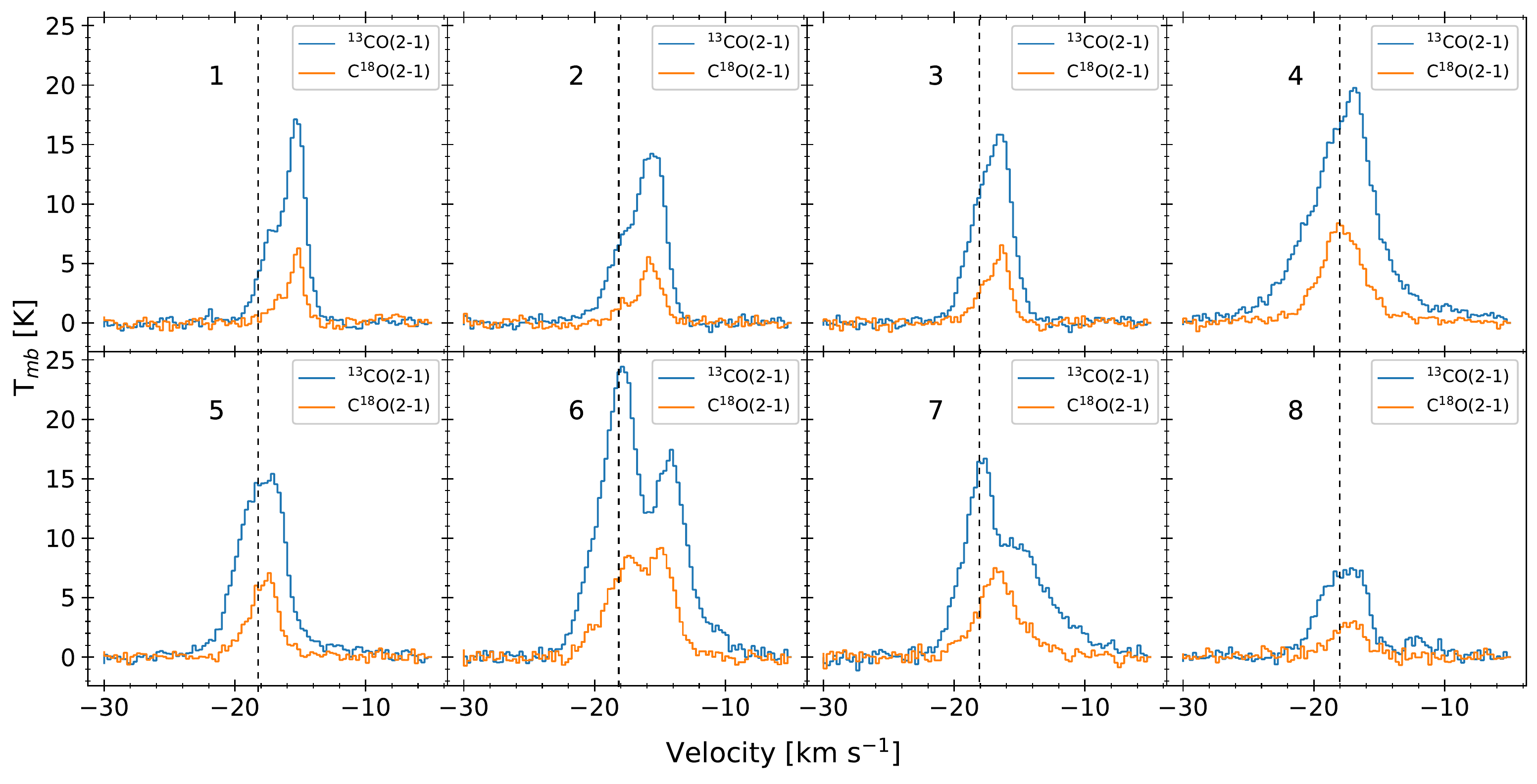}
\caption{The $^{13}$CO(J = 2--1) (in blue) and C$^{18}$O(J = 2--1) (in orange) spectra toward eight areas as highlighted by circles 
in Figure~\ref{fig7}f. A vertical line at V$_\mathrm{lsr}$ = $-$18 km s$^{-1}$ is also indicated in each panel.} 
\label{fig8}
\end{figure*}

Figure~\ref{fig8} shows the averaged spectra of $^{13}$CO(J = 2--1) and C$^{18}$O(J = 2--1) over eight small circles (radius = 20$''$)
distributed toward the elongated molecular cloud (see Figure~\ref{fig7}f). 
For a reference purpose, a vertical dashed line at V$_\mathrm{lsr}$ = $-$18 km s$^{-1}$ is also marked in each panel of Figure~\ref{fig8}. 
The circle nos. \#1--3 are distributed toward the bubble, while the circle nos. \#4 and \#6 are located toward the sites i17008 and i17009, respectively. The spectra of $^{13}$CO(J = 2--1) and C$^{18}$O(J = 2--1) show a single velocity peak toward four positions (\#3, 4, 5,  and 8).
However, we may see at least two velocity peaks in the direction of the other four positions (\#1, 2, 6, and 7). 
In particular, based on the $^{13}$CO(J = 2--1) and C$^{18}$O(J = 2--1) spectra toward the circle \#6, two velocity peaks are clearly found, allowing us to identify two cloud components at [$-$15.25, $-$11] km s$^{-1}$ (around $-$15 km s$^{-1}$) and [$-$22.25, $-$16] km s$^{-1}$ (around $-$18 km s$^{-1}$). 
Furthermore, with respect to the reference line, we find a change in the velocity peaks of molecular spectra on moving from 
the northern to southern parts of the cloud, showing the existence of a velocity gradient in the molecular cloud.

\begin{figure*}
\center
\includegraphics[width=16cm]{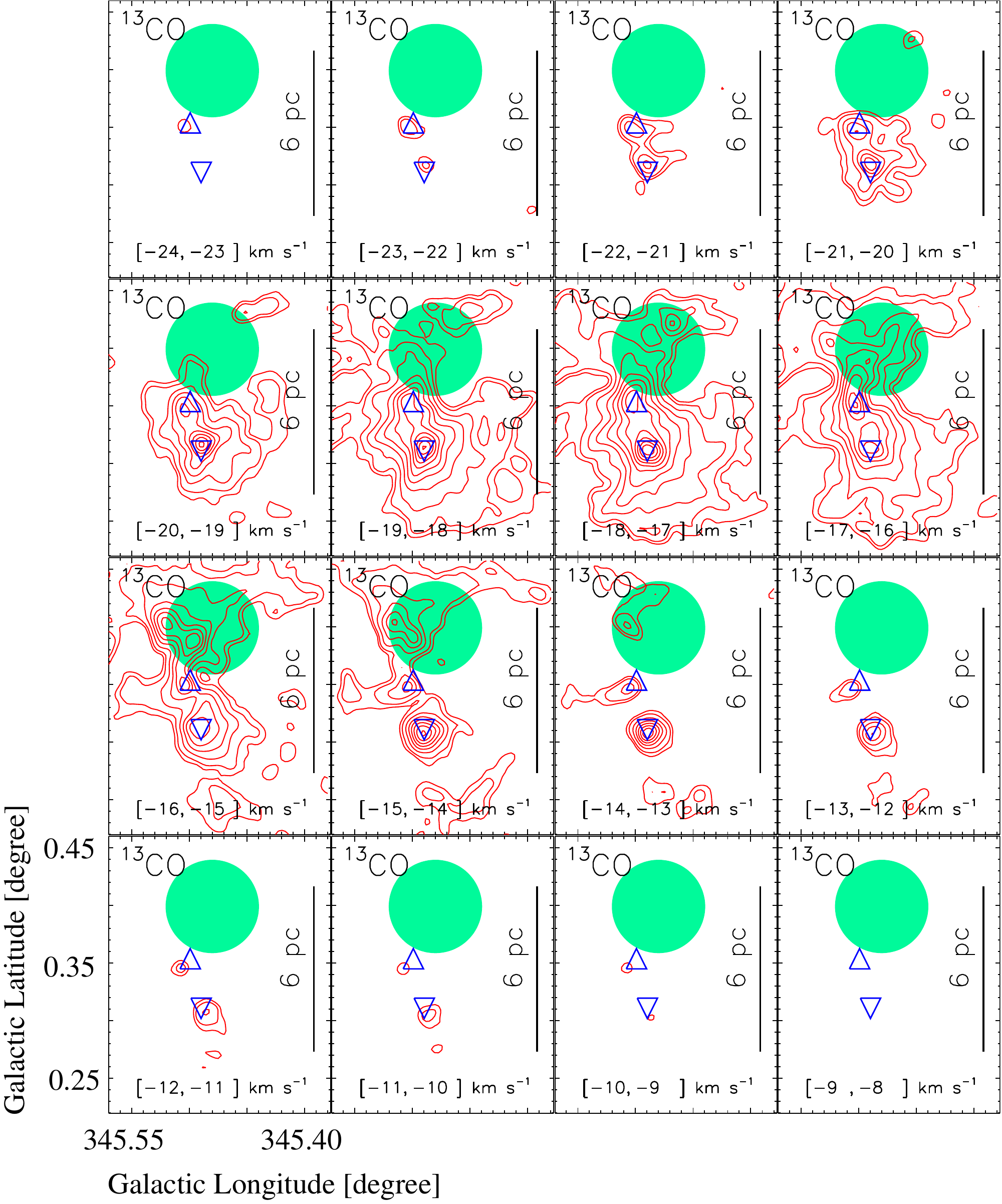}
\caption{Integrated velocity channel contours (in red) of $^{13}$CO(J = 2--1) emission (at velocity intervals of 1 km s$^{-1}$) 
toward an area containing i17008 and i17009. 
The contour levels are 2.4, 3.5, 6, 9, 12, 15, 18, 21, 24, 27, and 30 K km s$^{-1}$. 
The scale bar and other symbols are the same as in Figure~\ref{fig7}a. 
In each panel, a filled circle shows the location of the MIR bubble S11.}
\label{fig9}
\end{figure*}

\begin{figure*} 
\center
\includegraphics[width=\textwidth]{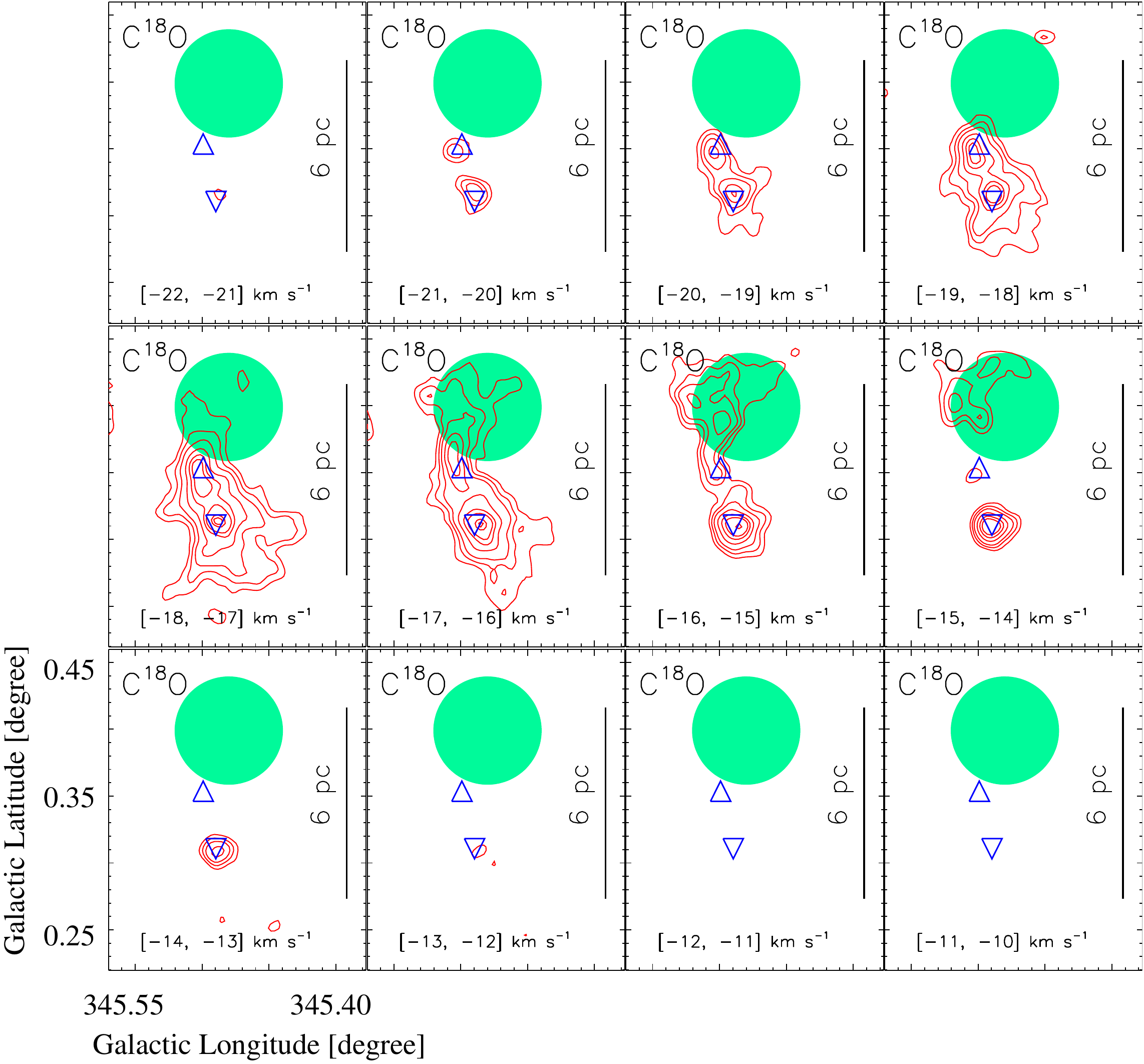}
\caption{Integrated velocity channel contours (in red) of C$^{18}$O(J = 2--1) emission (at velocity intervals of 1 km s$^{-1}$) 
toward an area containing i17008 and i17009. 
The contour levels are 1.5, 2.4, 3.5, 5.5, 7.5, 9, 12, and 13 K km s$^{-1}$. 
The scale bar and other symbols are the same as in Figure~\ref{fig7}a. 
In each panel, a filled circle shows the location of the MIR bubble S11.} 
\label{fig10}
\end{figure*}

Figures~\ref{fig9} and~\ref{fig10} display the integrated velocity channel contours of $^{13}$CO(J = 2--1) and C$^{18}$O(J = 2--1) (at velocity intervals of 1 km s$^{-1}$), respectively. Both channel maps support the presence of two molecular cloud components toward the area hosting two IRAS sources and the bubble 
(see panels at [$-$19, $-$18] and [$-$16, $-$15] km s$^{-1}$). The channel maps also support the presence of the HFS toward each IRAS site (see panels between [$-$22, $-$21] and [$-$17, $-$16] km s$^{-1}$). 

\begin{figure*}
\center
\includegraphics[width=\textwidth]{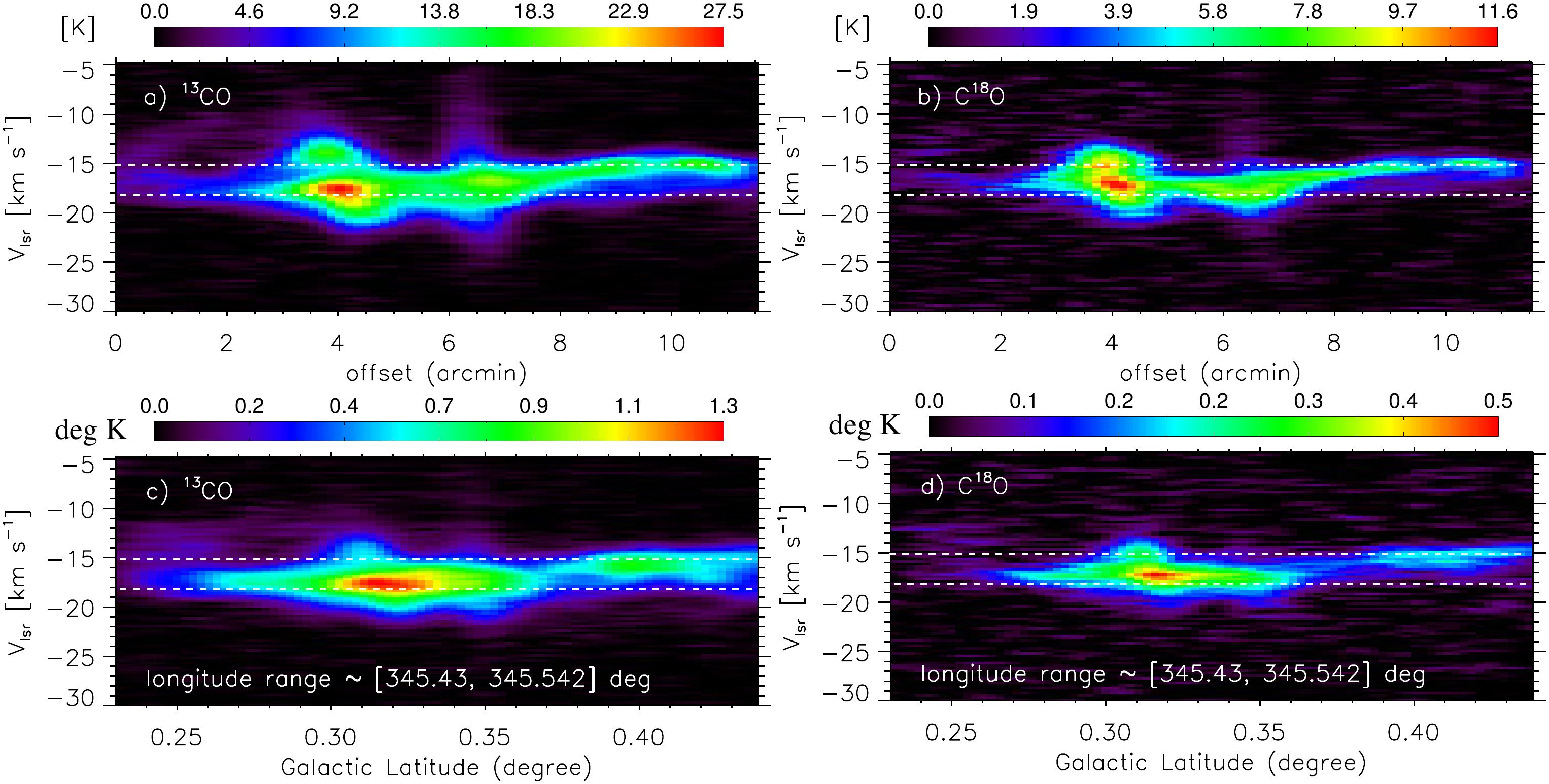}
\caption{Position-velocity diagram of a) $^{13}$CO(J = 2--1); b) C$^{18}$O(J = 2--1) along the 
curve highlighted in Figures~\ref{fig7}a and~\ref{fig7}d.
Latitude-velocity diagram of c) $^{13}$CO(J = 2--1); d) C$^{18}$O(J = 2--1). 
The molecular emission is integrated over a longitude range, 
which is indicated in panels ``c" and ``d". 
In all panels, the horizontal dashed lines (in white) are shown at V$_\mathrm{lsr}$ = $-$18 and $-$15 km s$^{-1}$.}
\label{fig11}
\end{figure*}

Figures~\ref{fig11}a and~\ref{fig11}b show the position-velocity diagrams of $^{13}$CO(J = 2--1) and C$^{18}$O(J = 2--1) along the curve as marked in Figures~\ref{fig7}a and~\ref{fig7}d, respectively. 
In Figures~\ref{fig11}c and~\ref{fig11}d, we present the latitude-velocity diagrams of $^{13}$CO(J = 2--1) and C$^{18}$O(J = 2--1) for a longitude range of 345$\degr$.43 to 345$\degr$.54. A continuous velocity structure is seen in all the position-velocity diagrams. These maps also suggest the presence of an outflow toward i17008 
and the presence of two velocity components (around $-$15 and $-$18 km s$^{-1}$). 
All these results together favour the spatial and velocity connections of two cloud components in the direction of i17008, i17009, and S11.

Figures~\ref{fig12}a and~\ref{fig12}b present the overlay of the $^{13}$CO(J = 2--1) and C$^{18}$O(J = 2--1) emission contours of two cloud components (at [$-$15.25, $-$11] and [$-$22.25, $-$16] km s$^{-1}$) on the {\it Spitzer} 8.0 $\mu$m image, respectively. The location of the elongated filament is also indicated in Figures~\ref{fig12}a and~\ref{fig12}b. 
In the {\it Spitzer} 8.0 $\mu$m image, the extended emissions detected toward both the IRAS sites and the bubble (or ionized clumps ``A--C") are seen toward the overlapping areas of two clouds. 

On the basis of the channel maps, we produce an integrated emission map of $^{13}$CO(J = 2--1) at [$-$17.75, $-$16.5] km s$^{-1}$ to trace the proposed HFSs (see Figure~\ref{fig12}c). In this relation, we exposed this $^{13}$CO(J = 2--1) map to an edge detection algorithm \citep[i.e. Difference of Gaussian (DoG); see][]{gonzalez02,assirati14,dewangan17b}.
In Figure~\ref{fig12}d, we display a two-color composite map made using the $^{13}$CO(J = 2--1) maps, which consists of the ``Edge-DoG" processed $^{13}$CO(J = 2--1) map at [$-$17.75, $-$16.5] km s$^{-1}$ (in red) and the $^{13}$CO(J = 2--1) map at [$-$24, $-$9] km s$^{-1}$ (in turquoise). 
In Figure~\ref{fig12}d, at least five curves are marked and labeled in the direction of i17008 (or ionized clump ``B"), 
while at least three curves are highlighted toward i17009 (or ionized clump ``C"). 
In Figure~\ref{fig12}d, one can clearly see the HFS toward each IRAS site. 

In Figure~\ref{fig12}e, we present the integrated emission map (moment-0) at [$-$21, $-$14] km s$^{-1}$ of the dense gas tracer N$_{2}$H$^{+}$(1--0) from the MALT90 data sets, which were observed for the areas covering G345.504/i17008 and G345.487/i17009 (see dotted-dashed boxes in Figure~\ref{fig12}c). Note that the MALT90 line data sets are not available toward the bubble S11.
The location of the elongated filament is also highlighted in the moment-0 map.  
The moment-0 map of N$_{2}$H$^{+}$ clearly displays the filamentary morphology as seen in the ATLASGAL continuum map at 870 $\mu$m. 

Figure~\ref{fig12}f displays the position-velocity diagram of N$_{2}$H$^{+}$ data set along the axis as marked in the N$_{2}$H$^{+}$ map (see Figure~\ref{fig12}e). In the position-velocity diagram of the N$_{2}$H$^{+}$ line, the hyperfine components are also evident. The position-velocity diagram supports the existence of a continuous velocity structure along the selected axis or filament hosting G345.504/i17008 and G345.487/i17009, and also shows a velocity spread toward the site i17009. 

The implication of all these observed findings is presented in Section~\ref{sec:disc}.
\section{Discussion}
\label{sec:disc}
In the direction of our selected target field around {\it l} = 345$\degr$.5, this paper mainly focuses on the elongated filamentary structures traced in different emissions (i.e., dust, molecular, and ionized). 
One of the new findings of this work is the presence of two distinct ionized filaments (i.e., IF-A and IF-B) located at different 
distances (see Section~\ref{sec:morph2}). 
Interestingly, the parent molecular clouds of both the ionized filaments are depicted in the same velocity range of [$-$21, $-$10] km s$^{-1}$, and have filamentary appearances. Several ionized clumps, which are excited by massive stars, are depicted toward the ionized filaments (see Figure~\ref{fig1} and Section~\ref{sec:morph}). 
In the following sections, we discuss the origin of massive stars and elongated ionized structures. 
\subsection{Interacting filamentary molecular clouds}
\label{sec:zzfffx}
We investigate at least two cloud components toward the parent molecular clouds of both the ionized filaments (see Section~\ref{sec:gasmorphb}). 
It is another new finding of this work. In the direction of each parent molecular cloud, we find a velocity connection of two cloud components having a velocity separation of about 3 km s$^{-1}$ (see Figure~\ref{fig6}). These filamentary clouds also spatially overlap with each other along the major axis, backing the existence of their multiple common zones. It may be considered as one of the forms of molecular/dust filamentary twisting/coupling \citep[e.g., LBN 140.07+01.64;][]{dewangan21}. 
This argument is valid for both the parent molecular clouds of IF-A and IF-B. 
These results together hint at the onset of the interaction or collision of filamentary molecular clouds, but do not favour a single point collision event of two molecular clouds. In the direction of the converging areas of the clouds, we find either dust clumps hosting massive stars or only ionized clumps powered by massive stars.

Earlier, signatures of the colliding flows were reported to Lupus~I \citep{gaczkowski15,gaczkowski17,krause18}. 
Lupus~I is associated with the Lupus clouds, which are nearby (150-200 pc) and young (1-2 Myr) star-forming region \citep[e.g.,][]{gaczkowski15}. 
Lupus~I is spatially seen between the Upper-Scorpius (USco) HI shell and the Upper Centaurus-Lupus (UCL) wind bubble \citep{gaczkowski15}. 
In other words, Lupus~I is thought to be situated along a filament at the converging area 
of these two bubbles, where the higher level of clumpiness is observed. 
In this context, Lupus~I has been suggested to be strongly influenced by colliding flows/shocked flows, which are produced by the 
expanding USco HI shell and the UCL wind bubble \citep{gaczkowski15,gaczkowski17,krause18}. 
These earlier works encourage us to explore the scenario of the colliding flows in our selected target area. 

Numerical simulations of the cloud cloud collision (CCC) process show the presence of massive and dense clumps/cores at the junction of two molecular clouds or the shock-compressed interface layer \citep[e.g.,][and references therein]{habe92,anathpindika10,inoue13,haworth15a,haworth15b,torii17,balfour17,bisbas17}, which is a very suitable environment for the MSF. In other words, massive stars and clusters of YSOs can be formed inside the dense gas layer produced via the strong compression at 
the colliding interface. 
In this relation, several observational works have been reported in the literature \citep[e.g.,][]{torii11,torii15,torii17,fukui14,fukui18,fukui21,dhanya21}.
Observationally, in the CCC process, one may expect a bridge feature in position-velocity diagrams, showing a connection of two clouds by an intermediate velocity and low intensity feature in velocity space \citep[e.g.,][]{haworth15b,dewangan17s235,dewangan18b,Kohno18,Priestley21}.
In addition, one may also expect a complementary distribution (i.e., a spatial match of ``key/intensity-enhancement" and ``cavity/keyhole/intensity-depression" features) in the collision event \citep[e.g.,][]{fukui18,dewangan18N36,Enokiya21}. However, we do not find any complementary distribution of two clouds in our target area.

\begin{figure}
\center
\includegraphics[width=8 cm]{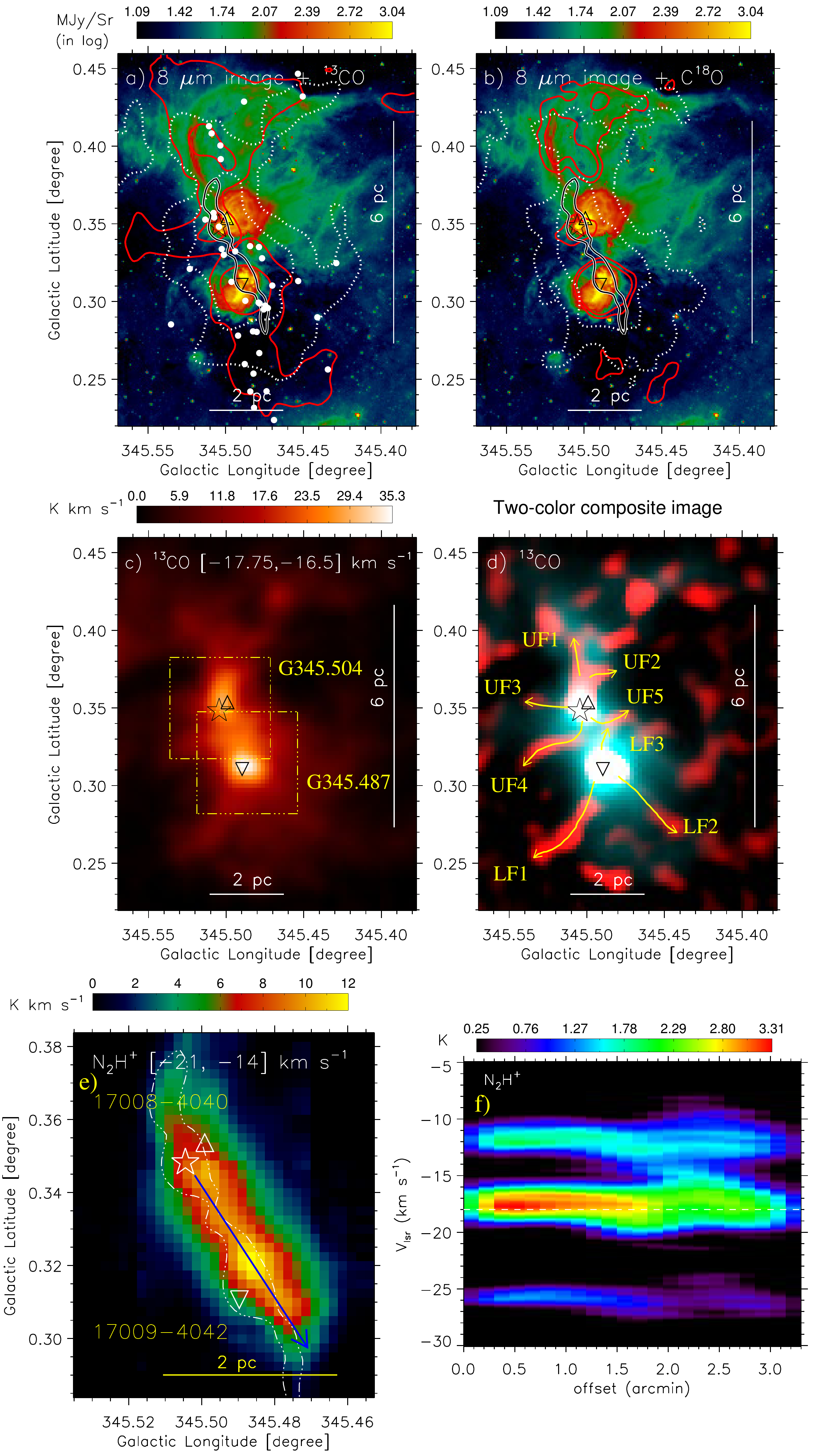}
\caption{Different cloud components toward an area containing i17008 and i17009.
a) Overlay of the $^{13}$CO(J = 2--1) emission of two clouds and the positions of previously reported YSOs \citep[see filled white circles; from][]{dewangan18} on the {\it Spitzer} 8.0 $\mu$m image. 
The molecular emission at [$-$15.25, $-$11] km s$^{-1}$ is shown by solid red contours with the levels of [5.35, 15.86] K km s$^{-1}$.
The dotted contours (in white) are the molecular emission at [$-$22.25, $-$16] km s$^{-1}$ with the levels of [14, 33.16] K km s$^{-1}$. 
b) Overlay of the C$^{18}$O(J = 2--1) emission of two clouds on the {\it Spitzer} 8.0 $\mu$m image. 
The molecular emission at [$-$15.25, $-$11] km s$^{-1}$ is shown by solid red contours with the levels of [2.2, 19.42] K km s$^{-1}$.
The dotted contours (in white) are the molecular emission at [$-$22.25, $-$16] km s$^{-1}$ with the levels of [2.5, 41.72] K km s$^{-1}$. 
c) The intensity map of $^{13}$CO(J = 2--1) integrated over a velocity range of [$-$17.75, $-$16.5] km s$^{-1}$. 
The areas covered by the MALT90 survey toward G345.504/i17008 and G345.487/i17009 are indicated by dotted-dashed boxes (see also Figure~\ref{fig12}e).
d) The panel displays a two-color composite map made using the $^{13}$CO(J = 2--1) maps. 
In the color-composite image, the ``Edge-DoG" processed $^{13}$CO(J = 2--1) map at [$-$17.75, $-$16.5] km s$^{-1}$ is presented in red color, 
while the $^{13}$CO(J = 2--1) map at [$-$24, $-$9] km s$^{-1}$ is shown in turquoise color. 
Based on visual inspection, eight curves are marked and labeled in the panel. 
e) Intensity map of the molecule N$_{2}$H$^{+}$(1--0) (from MALT90 data set) integrated 
over a velocity range of [$-$21, $-$14] km s$^{-1}$. 
f) Position-velocity diagram of N$_{2}$H$^{+}$(1--0) along an arrow as highlighted in the N$_{2}$H$^{+}$(1--0) map (see Figure~\ref{fig12}e). The position-velocity map of N$_{2}$H$^{+}$ was smoothed with a Gaussian function with a width of 3 pixels. A horizontal line at V$_\mathrm{lsr}$ = $-$18 km s$^{-1}$ is also indicated in the panel. 
In each panel (except panel ``f"), the scale bar and other symbols are the same as in Figure~\ref{fig7}a. 
In panels ``a" and ``b", a black curve highlights the elongated filament.} 
\label{fig12}
\end{figure}

Our results enable us to propose the applicability of the collision of two filamentary clouds in areas hosting IF-A and IF-B. 
Previously, in a massive-star forming region S237, a cluster of YSOs and a massive clump were found toward 
at the intersection of filamentary features, and the collision of these features was proposed to explain the observed cluster formation \citep{dewangan17a}. \citet{dewangan19} also identified two closely spaced (in velocity as well as in position) filamentary clouds in star-forming site AFGL 5142 and deciphered the presence of young stellar clusters by the filamentary collision/interaction scenario. 
In the case of the AFGL 333-Ridge, \citet{liang21} traced two velocity components having a velocity separation of about 2.5 km s$^{-1}$. 
Based on the analysis of $^{13}$CO line data, they proposed a scenario of colliding and merging of two cloud components into one molecular cloud in the AFGL 333-Ridge. 

In the present study, we consider the spatial extent of the overlapping regions of the two clouds, having a velocity separation of $\sim$3 km s$^{-1}$, to be $\sim$1.75 pc. 
We estimate the collision time-scale (i.e., the time-scale for which the material is accumulated at the collision zones) using the following equation \citep[see also][]{henshaw13} 
\begin{equation}
t_{\rm accum} = 2.0\,\bigg(\frac{l_{\rm fcs}}{0.5\,{\rm pc}} \bigg) \bigg(\frac{v_{\rm
rel}}{5{\rm \,km\,s^{-1}}}\bigg)^{-1}\bigg(\frac{n_{\rm pstc}/n_{\rm
prec}}{10}\bigg)\,{\rm Myr} 
\end{equation}
where, $n_{\rm prec}$ and $n_{\rm pstc}$ are the mean densities of the pre-collision and post-collision region, respectively.
Here, $l_{\rm fcs}$ is the collision length-scale and ${v_{\rm rel}}$ is the observed relative velocity.
In the present case, we do not know the exact viewing angle of the collision.
Therefore, a typical viewing angle of 45$\degr$ results in the collision length-scale ($l_{\rm fcs}$) of $\sim$2.5 pc (= 1.75 pc/sin(45$\degr$)), and the observed relative velocity (${v_{\rm rel}}$) of $\sim$4.2 km s$^{-1}$ (= 3 km s$^{-1}$/cos(45$\degr$)).
In this work, we do not have reliable estimates of $n_{\rm prec}$ and $n_{\rm pstc}$ values. 
However, logically, we expect $n_{\rm pstc}$ $>$ $n_{\rm prec}$ in a collision process, resulting in the higher mean density ratio ($\geq$1) of the post- and pre-collision regions. Considering a wide range of the mean density ratio of 1--10, we compute a range of collision timescale of $\sim$1.2--11.7 Myr.
It implies that the collision of two clouds might have occurred $\sim$1.2 Myr ago. 

In Section~\ref{sec:morph}, the dynamical ages of the ionized clumps (``A--G") are computed to be $\sim$0.1--1 Myr.
In the direction of site i17008, an O-star candidate without an H\,{\sc ii} region has also been investigated. 
Also, the noticeable Class~I protostars (mean age $\sim$0.44 Myr) appear to be seen toward the parent clouds of IF-A and IF-B (see Figure~\ref{fig2}d).   
Thus, considering different ages concerning signposts of star formation activities, we notice that the collision timescale is old enough to influence the star formation (including massive stars) in the parent molecular clouds of both the ionized filaments. Therefore, the star formation history in our target area seems to be explained by the collision of the two filamentary clouds.

The previously reported HFSs toward i17008 and i17009 are spatially seen in one of the cloud components (i.e., around $-$18 km s$^{-1}$; see Figure~\ref{fig12}).
The presence of HFSs may favour the onset of the global non-isotropic collapse (GNIC) scenario \citep[see][for more details]{Tige+2017,Motte+2018}. 
In the smoothed particle hydrodynamics simulations related to head-on collision of two clouds, \citet{balfour15} reported the presence of a pattern of filaments (e.g., hub or spokes systems) as resultant from the collision process. 
The theoretical work supports the origin of a shock-compressed layer by the colliding clouds, which fragments into filaments. Then these filaments form a network like a spider's web in the case of higher relative velocity between clouds (see also magnetohydrodynamic (MHD) simulations of \citet{inoue18}).  
Recently, the review article on CCC of \citet{fukui21} also stated that the onset of the collision process can produce 
hub filaments with their complex morphology. Using the SEDIGISM $^{13}$CO and C$^{18}$O line data, in the filamentary infrared dark cloud (IRDC) G333.73+0.37, \citet{dewangan2022new} presented the results in favour of CCC or converging flows, explaining the presence of the HFS and massive stars in the IRDC. Using the N$_{2}$H$^{+}$(1--0) observations, \citet{beltran22} explained the formation of the HFS and the origin of massive protocluster associated with the hot molecular core in the G31.41+0.31 cloud through the CCC. In the case of N159E-Papillon Nebula located in the Large Magellanic Cloud (distance $\sim$50 kpc), \citet{fukui19ex} provided observational results to support the scenario of the large-scale colliding flow, which was used to explain the existence of of massive stars and HFSs. These observational findings strongly support the connection of the formation of HFSs and massive stars with the CCC. Hence, our proposed collision process may also explain the existence of hubs in our target area. 

Overall, the interaction of elongated molecular filaments seems to be responsible for the 
birth of massive stars associated with IF-A and IF-B. 
\subsection{Ionized filaments IF-A and IF-B}
\label{sec:fffx}
In recent years, a wealth of studies on dust and molecular filaments have been carried out in star-forming sites, which strongly 
support their key role in the formation of stellar clusters and massive stars. 
Despite the availability of numerous radio continuum surveys, so far a very limited number of studies have conducted to investigate elongated ionized structures with high aspect ratios (length/thickness) in massive star-forming regions (e.g., Lynds Bright Nebulae \citep{karr03}, Eridanus filaments \citep{pon14}, Cygnus~X \citep{emig22}), which can be referred to as ionized filaments. 

In general, the study of massive star-forming regions is largely focused on H\,{\sc ii} regions powered by massive OB stars, which are often surrounded by MIR bubbles having different morphologies \citep[i.e., a complete or closed ring, a broken or incomplete ring, and a bipolar structure;][]{churchwell06}. In this context, one may not expect a very large elongated morphology of a single H\,{\sc ii} region excited by massive OB stars. 
In the literature, the ionized nature of the Lynds Bright Nebulae was explained by ultraviolet photons leaking from the nearby star-forming region W5 \citep{karr03}. In the case of Eridanus filaments, \citet{pon14} reported that these filaments are 
non-equilibrium structures, and might have produced when the Orion-Eridanus superbubble compressed a pre-existing gas cloud and swept up the gas into a dense ring around the outer boundary of the bubble. 
In the site Cygnus~X, \citet{emig22} proposed that the energetic feedback from Cyg OB2 (i.e., ultraviolet (UV) radiation, stellar winds, and radiation pressure) may be responsible for the observed ionized filaments via swept-up ionized gas or dissipated turbulence. 
Hence, the common explanation of the origin of ionized filaments is likely due to the feedback from massive stars. 

Most recently, \citet{whitworth21} studied a semi-analytic model concerning ionizing feedback from an O star formed in a filament. 
According to the model, the filament is generally destroyed by the ionizing radiation from the O star, 
and the ionized gas disperses freely into the surroundings. We refer to this as a ``case-I" phase in this work.
In the case of relatively wide and/or relatively dense filament and/or low the rate at which the O star emits ionizing photons, 
the ongoing accretion inflow on to the filament will reduce 
the escape of ionized gas, and might trap the ionizing radiation from the O star. 
This will slow the erosion of the filament, and the model also shows the formation of a relatively dense, compact, and turbulent
H\,{\sc ii} region around the ionizing stars. We refer to this as a ``case-II" phase.

The spatial association of molecular gas with the ionized structures seems to indicate that the filamentary molecular clouds are pre-existing structures (see Figures~\ref{fig2} and~\ref{fig4}). 
Hence, the filamentary molecular clouds are unlikely to be formed by the feedback of young massive OB 
stars associated with the ionized filaments. 
However, the energetics of massive stars appear to have influenced their parent filamentary molecular clouds (see Figures~\ref{fig2} and~\ref{fig4}). 
In other words, massive stars wreak havoc on the gas and dust in the parental filamentary molecular clouds. 

The knowledge of three pressure components (i.e., pressure of an H\,{\sc ii} region ($P_{HII}$), radiation pressure (P$_{rad}$), and stellar wind ram pressure (P$_{wind}$)) driven by a massive OB star can be useful to explore the feedback of a massive star in its vicinity \citep[e.g.,][]{dewangan16}. 
The equations of different pressure components are $P_{HII} = \mu m_{H} c_{s}^2\, \left(\sqrt{3N_\mathrm{UV}\over 4\pi\,\alpha_{B}\, D_{s}^3}\right)$; P$_{rad}$ = $L_{bol}/ 4\pi c D_{s}^2$; and P$_{wind}$ = $\dot{M}_{w} V_{w} / 4 \pi D_{s}^2$ \citep[see][for more details]{bressert12,dewangan16}. 
In these equations, $N_\mathrm{UV}$ is defined earlier, c$_{s}$ is the sound speed of the photo-ionized gas \citep[i.e., 11 km s$^{-1}$;][]{bisbas09}, $\alpha_{B}$ is the radiative recombination coefficient \citep[=  2.6 $\times$ 10$^{-13}$ $\times$ (10$^{4}$ K/T$_{e}$)$^{0.7}$ cm$^{3}$ s$^{-1}$; see][]{kwan97}, $\mu$ is the mean molecular weight in the ionized gas
\citep[i.e., 0.678;][]{bisbas09}, m$_{H}$ is the hydrogen atom mass, $\dot{M}_{w}$ is the mass-loss rate, 
V$_{w}$ is the wind velocity of the ionizing source, 
L$_{bol}$ is the bolometric luminosity of the source, and D$_{s}$ is the projected distance from the position of a massive star where the pressure components are determined. 

In the case of Wolf-Rayet stars, the value of P$_{wind}$ dominates over the values of $P_{HII}$ and P$_{rad}$ \citep[e.g.,][]{lamers99,dewangan16xs,baug19}. But, the value of $P_{HII}$ driven by massive zero age main sequence stars often exceeds their P$_{rad}$ and P$_{wind}$ components \citep[e.g.,][]{dewangan16}. Hence, we have computed only the values of $P_{HII}$ in the direction of IF-A and IF-B. 
We find a total of N$_{uv}$ = 1.72 $\times$ 10$^{48}$ s$^{-1}$ of three clumps ``A--C'' in IF-A, while a total of N$_{uv}$ = 2.65 $\times$ 10$^{48}$ s$^{-1}$ of three clumps ``E--G'' in IF-B are estimated. Considering the elongated appearance of the filaments, we choose a value of D$_{s}$ = 3 pc for the calculations. Using $\alpha_{B}$ = 2.6 $\times$ 10$^{-13}$ cm$^{3}$ s$^{-1}$ at T$_{e}$ = 10$^{4}$~K and D$_{s}$ = 3 pc, we calculated the values of P$_{HII}$ to be $\approx$6.1 $\times$ 10$^{-11}$ and $\approx$7.6 $\times$ 10$^{-11}$ dynes\, cm$^{-2}$ for IF-A and IF-B, respectively. 
Each value can be compared with the pressure exerted by the self-gravity of the surrounding molecular gas around respective ionized filament.  Based on the detection of various molecular emission (e.g., C$^{18}$O (critical density $\sim$10$^{4}$ cm$^{-3}$); see Figures~\ref{fig3} and~\ref{fig5}), we assume the values of particle density $\geq$ 10$^{4}$ cm$^{-3}$ for IF-A, and $<$ 10$^{3}$ cm$^{-3}$ for IF-B. 
From Table 7.3 in \citet{dyson97}, we find pressure values (P$_{MC}$) for typical cool molecular clouds (particle density $\sim$ 10$^{3}$ -- 10$^{4}$ cm$^{-3}$ and temperature $\sim$ 20 K) to be $\sim$2.8 $\times$ 10$^{-12}$ -- 2.8 $\times$ 10$^{-11}$ dynes cm$^{-2}$. 

Concerning the parent molecular cloud of IF-A, an H\,{\sc ii} region powered by a massive star is traced 
at its three different locations. In other words, there are three H\,{\sc ii} regions toward the parent 
molecular cloud of IF-A. The combined feedback of three H\,{\sc ii} regions located in 
the elongated molecular cloud has not eroded the parent cloud, and may be responsible for the elongated ionized morphology of IF-A. 
It is supported by the pressure values (i.e., P$_{HII}$ $\approx$ P$_{MC}$) inferred toward IF-A.   
On the basis of the C$^{18}$O emission and the dust continuum emission, we consider the parent molecular cloud of IF-A as a dense filament. Therefore, the filamentary cloud associated with IF-A resembles the ``case-II" phase as described by \citet{whitworth21}. 

In a similar fashion, the existence of IF-B may be explained by the combined feedback of massive stars 
powering the ionized clumps ``D--G".
Furthermore, the central part of the parent molecular cloud of IF-B hosts bubble-like structures (with $T_\mathrm{d}$ = 21--27~K) as seen in the {\it Herschel} temperature map, and seems to be destroyed by the impact of the H\,{\sc ii} regions in the ionized filament IF-B. In this relation, one may also obtain the hint from the pressure values (i.e., P$_{HII}$ $>$ P$_{MC}$) in IF-B. After the erosion of the filamentary molecular cloud, the ionizing radiation has freely streamed out, 
which could lead to the elongated ionized morphology of IF-B. This implies the applicability of the ``case-I" phase as reported by \citet{whitworth21}. 
\subsection{Velocity structure function toward IF-A}
\label{xxssec:data3}
It is possible that the colliding filaments and the stellar feedback from massive stars may drive internal turbulence in the parent clouds of both the ionized filaments. In this section, to examine the properties of turbulence, we study the velocity structure function of a continuous elongated molecular structure associated with IF-A. 
In this connection, we determined the second-order structure function ($S_2(L)$) as reported in \citet{hacar16}, and 
the square root of the second-order structure function is defined as:
\begin{equation}
S_2(L)^{1/2} = \delta V =  \left<|V(r)-V(r+L)|^2\right>^{1/2} = \left<|\delta u_{l}|^2\right>^{1/2}
\label{xxshh1}
\end{equation}
Here, $\delta u_{l}$ = V(r) $-$ V(r+L) is the velocity difference between two positions separated by a lag. 
Based on the earlier reported work by \citet{heyer04}, we find that the velocity structure function defined in this way is a useful tool to study properties of turbulence in molecular clouds and can be directly compared with the Larson's size-linewidth relation \citep{larson81}. 

In this work, the structure functions are derived from the $^{13}$CO and N$_{2}$H$^{+}$ line data. 
The lines with different critical densities give an opportunity to study gas dynamics in less dense (extended) and more dense (compact) parts of the filament. Line velocities are determined from Gaussian fitting to the profiles.

\begin{figure*}
\includegraphics[width=11.5 cm]{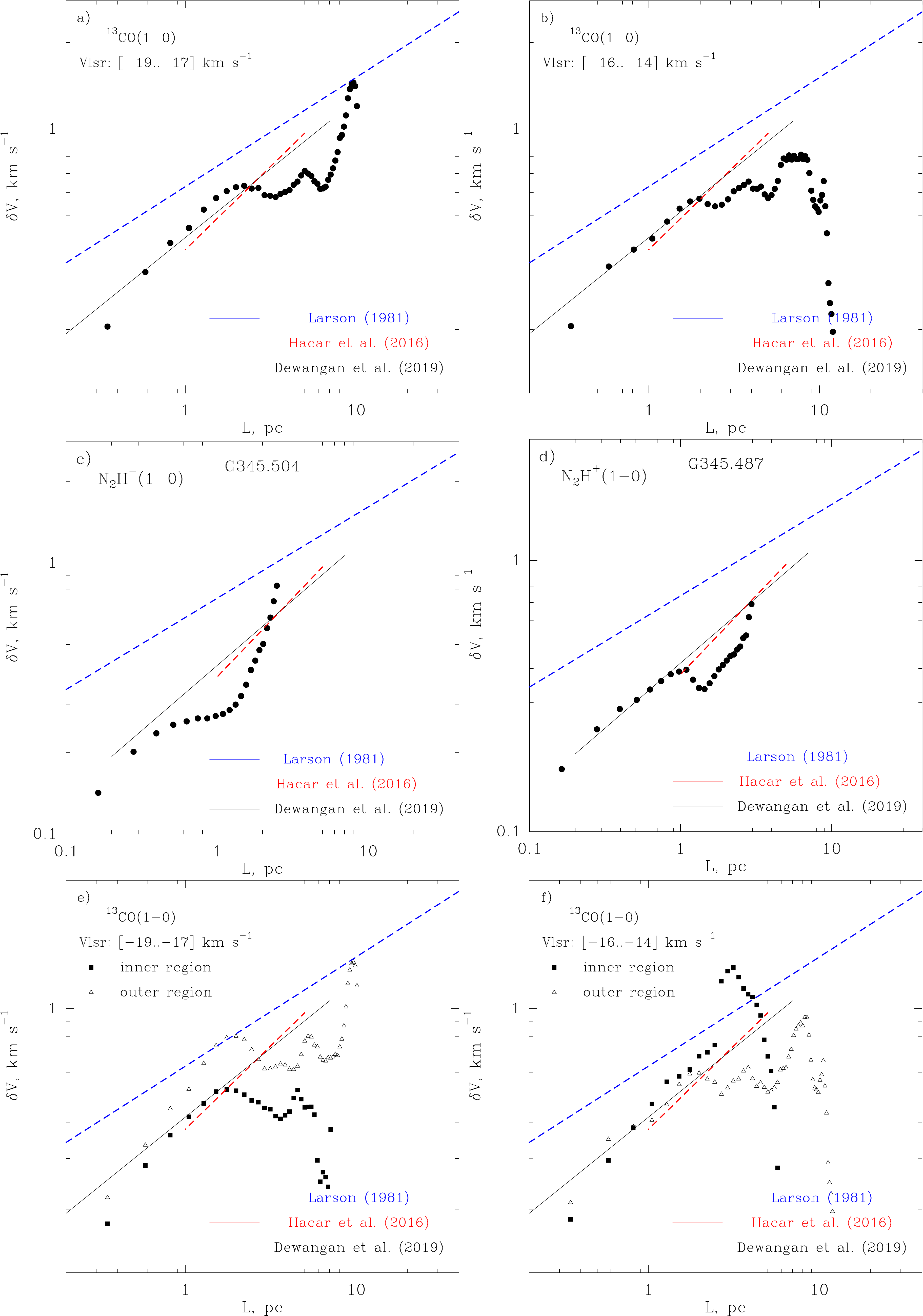}
\caption{All the panels show a structure function in velocity $\delta V$ as a function of length. 
a) The structure function is derived using the $^{13}$CO line data for a velocity range of [$-$19, $-$17] km s$^{-1}$ and 
for the total area containing the molecular filamentary region associated with IF-A (see Figure~\ref{fig7}a). 
b) The structure function is generated using the $^{13}$CO line data for a velocity range of [$-$16, $-$14] km s$^{-1}$ (see Figure~\ref{fig7}a). 
c) The structure function is produced using the N$_{2}$H$^{+}$ line data for an area related to G345.504 (see Figure~\ref{fig12}c). 
d) The structure function is derived using the N$_{2}$H$^{+}$ line data for an area related to G345.487 (see Figure~\ref{fig12}c). 
e) The structure function is computed using the $^{13}$CO line data for a velocity range of [$-$19, $-$17] km s$^{-1}$ and for an area located within (i.e., inner/filamentary region; see filled squares) and outside (outer region; see open triangles) of the $^{13}$CO contour as shown in Figure~\ref{fig7}a.
f) The structure function is computed using the $^{13}$CO line data for a velocity range of [$-$16, $-$14] km s$^{-1}$ 
and for inner regions (see filled squares) and outer regions (see open triangles) of the filament. 
In panels ``c" and ``d", line velocities lie within a velocity range of [$-$19, $-$17] km s$^{-1}$. 
In each panel, the Larson's velocity dispersion-size relationship \citep[i.e., $\delta V$ = 0.63 $\times$ L$^{0.38}$;][]{larson81} is marked by a dashed blue line. 
In all panels, a solid black line shows the relation, $\delta V$ = 0.42 $\times$ L$^{0.48}$ related to the star-forming site S242 \citep[from][]{dewangan19x}. In each panel, the velocity dispersion-size relationship of the Musca cloud is highlighted by a dashed red line \citep[e.g., $\delta V$ = 0.38 $\times$ L$^{0.58}$;][]{hacar16}. 
A lag corresponds to 20$''$ and 10$''$ for the areas using the $^{13}$CO and N$_{2}$H$^{+}$ line data, respectively.} 
\label{fig13}
\end{figure*}

Using the $^{13}$CO line data, structure functions are constructed for two different velocity ranges of [$-$19, $-$17] km s$^{-1}$ and [$-$16, $-$14] km s$^{-1}$ as presented in Figures~\ref{fig13}a and~\ref{fig13}b, respectively. In the calculations, the $^{13}$CO integrated intensities (I($^{13}$CO)) $>$ 3 K km s$^{-1}$ and a lag of 20$''$ are considered. A range of linewidth of [1, 3] km s$^{-1}$ is adopted in order to exclude the spectra with overlapping components in the analysis.

The N$_{2}$H$^{+}$ line data are known to trace dense gas in a given star-forming region. 
Therefore, to study structure functions toward the dense clumps hosting IRAS 17008-4040/G345.504 and IRAS 17009-4042/G345.487, 
the N$_{2}$H$^{+}$ line data are employed. Using the N$_{2}$H$^{+}$ line data, Figures~\ref{fig13}c and~\ref{fig13}d present structure functions for an area hosting IRAS 17008-4040/G345.504 and IRAS 17009-4042/G345.487, respectively. 
The calculations use a velocity range of [$-$19, $-$17] km s$^{-1}$, a lag of 10$''$, and a range of linewidth of [1, 3] km s$^{-1}$. 

In general, concerning the study of turbulence in molecular clouds, the Larson's one-dimensional velocity dispersion-size relationship or Larson's scaling relation \citep[i.e., $\delta V$ = 0.63 $\times$ L$^{0.38}$;][]{larson81} can be examined. 
In this relation, one can expect the dominant turbulent flow against the kinematically coupled large-scale, ordered motion in molecular clouds. Using the $^{13}$CO line data, \citet{hacar16} studied the velocity structure function for the Musca cloud (i.e., $\delta V$ = 0.38 $\times$ L$^{0.58}$), 
which was found to be different from the Larson's scaling relation. This deviation was attributed due to the presence of sonic-like structure in the cloud which is decoupled with the dominant turbulent velocity structure. 

In Figures~\ref{fig13}a--\ref{fig13}d, we have also shown the Larson's scaling (i.e., $\delta V$ = 0.63 $\times$ L$^{0.38}$), the relation concerning the star-forming site S242 \citep[i.e., $\delta V$ = 0.42 $\times$ L$^{0.48}$;][]{dewangan19x}, 
and a relationship of the Musca cloud (i.e., $\delta V$ = 0.38 $\times$ L$^{0.58}$) by dashed blue line, solid black line, and dashed red line, respectively. 

Structure functions of the selected regions show nearly power-law dependencies for lower L ($\leq$ 2 pc using $^{13}$CO and $\leq$ 0.5--1 pc using N$_{2}$H$^{+}$; see the X-axis in Figure~\ref{fig13}). 
For higher L, structure functions behave mostly in an irregular manner. In Figures~\ref{fig13}a--\ref{fig13}d, all the structure functions as those for S242 and Musca lie under the Larson's dependence. The power-law dependencies of the structure functions using the $^{13}$CO [$-$16,$-$14] km s$^{-1}$ (see Figure~\ref{fig13}b) and for the region G345.487 using the N$_{2}$H$^{+}$ line data (see Figure~\ref{fig13}d) turn out to be close to the one, which is also found for S242 having a power-law index of about 0.48. 
A power-law index of the structure function derived from the $^{13}$CO [$-$19, $-$17] km s$^{-1}$ data is about 0.6. 
These indices are higher than of the Larson's dependence, and are predicted for supersonic incompressible turbulence with intermittency \citep[e.g.,][]{she94,boldyrev02}. In order to compare our estimated indices with the predictions of the highlighted papers, one has to multiply our indices by 2  as we used a square root of the structure function. Furthermore, using the $^{13}$CO line data, we also compared the structure functions derived for the inner subregion containing the filament against its surrounding (outer) gas. 
For the $^{13}$CO [$-$16, $-$14] km s$^{-1}$ range, no definite differences in the power-law indices are found. 
However, for the  $^{13}$CO [$-$19, $-$17] km s$^{-1}$ range, the index of the power-law dependence in surrounding gas is higher than for the filament (see Figures~\ref{fig13}e and~\ref{fig13}f). 

\citet{chira19} studied the evolution of simulated turbulent clouds under the influence of various physical processes and explored how velocity structure functions change with time. Their results show a general behavior of the structure function-L dependencies similar to ours (Figures~\ref{fig13}a--\ref{fig13}d) at early stages of the cloud evolution. The power-law indices of their structure function-L 
without density weighting -- which is our case -- appear to be higher than the theoretical predictions in accordance with our results. It is also found that the structure function's power-law indices decrease with time due to growing influence of systemic velocities in self-gravitating gas on small scales. 
In the direction of the ionized filament IF-A, different scales of the regions, where turbulence dominates as observed from the $^{13}$CO and N$_{2}$H$^{+}$ structure functions, are most likely due to the difference in critical densities of these molecular lines. 
The difference in the power-law indices of the structure function dependencies calculated using the $^{13}$CO [$-$19, $-$17] km s$^{-1}$ data for the region of the filament and for its surrounding gas could reflect the difference in turbulent properties connected with the 
feedback from massive stars in the filament. 

We conclude that the observed power-law dependencies of the structure functions derived from molecular line data for the ionized filament IF-A most probably reflect turbulent properties of gas. In order to confirm the properties of turbulence in massive star-forming cores, which differ from their surroundings, new observations with high sensitivity and high angular resolution using the lines with different critical densities are needed. Using the line data, the study of velocity structure functions, and their comparison with the theoretical predictions and the results of model simulations seems to be a useful tool for this purpose.

\section{Summary and Conclusions}
\label{sec:conc}
To observationally investigate the embedded morphology and ongoing physical mechanisms around {\it l} = 345$\degr$.5, we have carried an analysis of multi-wavelength data of an area of $\sim$74\rlap.{$'$}6 $\times$ 55$'$.  
The radio continuum map at 843 MHz reveals two distinct ionized filaments (i.e., IF-A (extent $\sim$8\rlap.{$'$}5) and IF-B (extent $\sim$22\rlap.{$'$}65)). Ionized clumps powered by massive OB stars are identified toward both the ionized filaments. 
The $^{13}$CO(1--0), $^{13}$CO(2--1), and C$^{18}$O(2--1) emissions are examined in a velocity range of [$-$21, $-$10] km s$^{-1}$ 
to study the parent molecular clouds of IF-A and IF-B, which have filamentary appearances. However, IF-A and IF-B seem to be situated at a distance of 2.4 kpc and 1.4 kpc, respectively. We have investigated two cloud components around $-$18 and $-$15 km s$^{-1}$ toward the filamentary parent clouds of IF-A and IF-B, which are connected in velocity space. The filamentary cloud components also spatially overlap with each other along the major axis, which may be treated as filamentary twisting/coupling. 
Massive stars are evident toward the common zones of the cloud components, where noticeable Class~I protostars also seem to be present.   
Based on our observational outcomes, we suggest the possibility of the collision of two filamentary clouds around 1.2 Myr ago. 
The origin of IF-A and IF-B may be explained by the combined feedback of massive stars. 

The continous elongated structure of the parent cloud of IF-A is identified in the molecular maps, and power-law dependencies of its structure functions are found, reflecting turbulent properties of gas. The difference in the power-law indices of the structure function dependencies for the gas in the filament and its surrounding gas is found. It could be connected with the influence of massive stars on the filament, which may affect the turbulent properties.

The central part of the parent cloud of IF-B is broken where this ionized filament is detected. Considering the observed ionized and molecular morphologies, our results seem to support the findings of the most recent model of Whitworth \& Priestley (2021), which is related to the escape and the trap of the ionizing radiation from an O star formed in a filament.
\section*{Acknowledgments}
We are grateful to the anonymous reviewer for the constructive comments and suggestions.   
The research work at Physical Research Laboratory is funded by the Department of Space, Government of India. L.E.P. acknowledges the support of the IAP State Program 0030-2021-0005. This work is based [in part] on observations made with the {\it Spitzer} Space Telescope, which is operated by the Jet Propulsion Laboratory, California Institute of Technology under a contract with NASA. 
This publication is based on data acquired with the Atacama Pathfinder Experiment (APEX) under programmes 092.F-9315 and 193.C-0584. APEX is a collaboration among the Max-Planck-Institut fur Radioastronomie, the European Southern Observatory, and the Onsala Space Observatory. The processed data products are available from the SEDIGISM survey database located at https://sedigism.mpifr-bonn.mpg.de/index.html, which was constructed by James Urquhart and hosted by the Max Planck Institute for Radio Astronomy. A part of this work has made use of data from the European Space Agency (ESA) mission Gaia, processed by the Gaia Data Processing and Analysis Consortium (DPAC). Funding for the DPAC is provided by national institutions, in particular, the institutions participating in the Gaia Multilateral Agreement. 
\subsection*{Data availability}
Distances to stars in the GAIA EDR3 underlying this article are available from the publicly accessible website\footnote[1]{https://cdsarc.cds.unistra.fr/viz-bin/cat/I/352}. 
The {\it Herschel}, WISE, and {\it Spitzer} data underlying this article are available from the publicly accessible NASA/IPAC infrared science archive\footnote[2]{https://irsa.ipac.caltech.edu/frontpage/}.
The {\it Herschel} temperature map underlying this article is available from the publicly accessible website\footnote[3]{http://www.astro.cardiff.ac.uk/research/ViaLactea/}.
The ATLASGAL 870 $\mu$m continuum data underlying this article are available from the publicly accessible ATLASGAL database server\footnote[4]{https://www3.mpifr-bonn.mpg.de/div/atlasgal/}.
The SEDIGISM molecular line data underlying this article are available from the publicly accessible website\footnote[5]{https://sedigism.mpifr-bonn.mpg.de/cgi-bin-seg/SEDIGISM\_DATABASE.cgi}.
The Mopra molecular line data underlying this article are available from the publicly accessible website\footnote[6]{https://dataverse.harvard.edu/dataset.xhtml?persistentId=doi:10.7910/DVN/LH3BDN}.
The MALT90 molecular line data underlying this article are available from the publicly accessible website\footnote[7]{http://malt90.bu.edu/}. 
The SUMSS 843 MHz continuum data underlying this article are available from the publicly accessible server\footnote[8]{https://skyview.gsfc.nasa.gov/current/cgi/query.pl}.
%-------------------------
%-- Figures
%\textwidth--------------------------------


\bibliographystyle{mnras}
\bibliography{reference} % if your bibtex file is called example.bib

\begin{thebibliography}{}
\makeatletter
\relax
\def\mn@urlcharsother{\let\do\@makeother \do\$\do\&\do\#\do\^\do\_\do\%\do\~}
\def\mn@doi{\begingroup\mn@urlcharsother \@ifnextchar [ {\mn@doi@}
  {\mn@doi@[]}}
\def\mn@doi@[#1]#2{\def\@tempa{#1}\ifx\@tempa\@empty \href
  {http://dx.doi.org/#2} {doi:#2}\else \href {http://dx.doi.org/#2} {#1}\fi
  \endgroup}
\def\mn@eprint#1#2{\mn@eprint@#1:#2::\@nil}
\def\mn@eprint@arXiv#1{\href {http://arxiv.org/abs/#1} {{\tt arXiv:#1}}}
\def\mn@eprint@dblp#1{\href {http://dblp.uni-trier.de/rec/bibtex/#1.xml}
  {dblp:#1}}
\def\mn@eprint@#1:#2:#3:#4\@nil{\def\@tempa {#1}\def\@tempb {#2}\def\@tempc
  {#3}\ifx \@tempc \@empty \let \@tempc \@tempb \let \@tempb \@tempa \fi \ifx
  \@tempb \@empty \def\@tempb {arXiv}\fi \@ifundefined
  {mn@eprint@\@tempb}{\@tempb:\@tempc}{\expandafter \expandafter \csname
  mn@eprint@\@tempb\endcsname \expandafter{\@tempc}}}

\bibitem[\protect\citeauthoryear{{Anathpindika}}{{Anathpindika}}{2010}]{anathpindika10}
{Anathpindika} S.~V.,  2010, \mn@doi [\mnras]
  {10.1111/j.1365-2966.2010.16541.x}, \href
  {https://ui.adsabs.harvard.edu/abs/2010MNRAS.405.1431A} {405, 1431}

\bibitem[\protect\citeauthoryear{{Andr{\'e}} et~al.,}{{Andr{\'e}}
  et~al.}{2010}]{andre10}
{Andr{\'e}} P.,  et~al., 2010, \mn@doi [\aap] {10.1051/0004-6361/201014666},
  \href {https://ui.adsabs.harvard.edu/abs/2010A&A...518L.102A} {518, L102}

\bibitem[\protect\citeauthoryear{{Andr{\'e}}, {Di Francesco}, {Ward-Thompson},
  {Inutsuka}, {Pudritz}  \& {Pineda}}{{Andr{\'e}} et~al.}{2014}]{andre14}
{Andr{\'e}} P.,  {Di Francesco} J.,  {Ward-Thompson} D.,  {Inutsuka} S.~I.,
  {Pudritz} R.~E.,   {Pineda} J.~E.,  2014, in {Beuther} H.,  {Klessen} R.~S.,
  {Dullemond} C.~P.,   {Henning} T.,  eds, Protostars and Planets VI. p.~27
  (\mn@eprint {arXiv} {1312.6232}),
  \mn@doi{10.2458/azu\_uapress\_9780816531240-ch002}

\bibitem[\protect\citeauthoryear{{Assirati}, {Silva}, {Berton}, {Lopes}  \&
  {Bruno}}{{Assirati} et~al.}{2014}]{assirati14}
{Assirati} L.,  {Silva} N.~R.,  {Berton} L.,  {Lopes} A.~A.,   {Bruno} O.~M.,
  2014, in Journal of Physics Conference Series. p. 012020 (\mn@eprint {arXiv}
  {1311.2561}), \mn@doi{10.1088/1742-6596/490/1/012020}

\bibitem[\protect\citeauthoryear{{Bailer-Jones}, {Rybizki}, {Fouesneau},
  {Demleitner}  \& {Andrae}}{{Bailer-Jones} et~al.}{2021}]{bailer21}
{Bailer-Jones} C.~A.~L.,  {Rybizki} J.,  {Fouesneau} M.,  {Demleitner} M.,
  {Andrae} R.,  2021, \mn@doi [\aj] {10.3847/1538-3881/abd806}, \href
  {https://ui.adsabs.harvard.edu/abs/2021AJ....161..147B} {161, 147}

\bibitem[\protect\citeauthoryear{{Balfour}, {Whitworth}, {Hubber}  \&
  {Jaffa}}{{Balfour} et~al.}{2015}]{balfour15}
{Balfour} S.~K.,  {Whitworth} A.~P.,  {Hubber} D.~A.,   {Jaffa} S.~E.,  2015,
  \mn@doi [\mnras] {10.1093/mnras/stv1772}, \href
  {https://ui.adsabs.harvard.edu/abs/2015MNRAS.453.2471B} {453, 2471}

\bibitem[\protect\citeauthoryear{{Balfour}, {Whitworth}  \& {Hubber}}{{Balfour}
  et~al.}{2017}]{balfour17}
{Balfour} S.~K.,  {Whitworth} A.~P.,   {Hubber} D.~A.,  2017, \mn@doi [\mnras]
  {10.1093/mnras/stw2956}, \href
  {https://ui.adsabs.harvard.edu/abs/2017MNRAS.465.3483B} {465, 3483}

\bibitem[\protect\citeauthoryear{{Baug}, {de Grijs}, {Dewangan}, {Herczeg},
  {Ojha}, {Wang}, {Deng}  \& {Bhatt}}{{Baug} et~al.}{2019}]{baug19}
{Baug} T.,  {de Grijs} R.,  {Dewangan} L.~K.,  {Herczeg} G.~J.,  {Ojha} D.~K.,
  {Wang} K.,  {Deng} L.,   {Bhatt} B.~C.,  2019, \mn@doi [\apj]
  {10.3847/1538-4357/ab46be}, \href
  {https://ui.adsabs.harvard.edu/abs/2019ApJ...885...68B} {885, 68}

\bibitem[\protect\citeauthoryear{{Beltr{\'a}n}, {Rivilla}, {Kumar}, {Cesaroni}
  \& {Galli}}{{Beltr{\'a}n} et~al.}{2022}]{beltran22}
{Beltr{\'a}n} M.~T.,  {Rivilla} V.~M.,  {Kumar} M.~S.~N.,  {Cesaroni} R.,
  {Galli} D.,  2022, \mn@doi [\aap] {10.1051/0004-6361/202243361}, \href
  {https://ui.adsabs.harvard.edu/abs/2022A&A...660L...4B} {660, L4}

\bibitem[\protect\citeauthoryear{{Benjamin} et~al.,}{{Benjamin}
  et~al.}{2003}]{benjamin03}
{Benjamin} R.~A.,  et~al., 2003, \mn@doi [\pasp] {10.1086/376696}, \href
  {https://ui.adsabs.harvard.edu/abs/2003PASP..115..953B} {115, 953}

\bibitem[\protect\citeauthoryear{{Bisbas}, {W{\"u}nsch}, {Whitworth}  \&
  {Hubber}}{{Bisbas} et~al.}{2009}]{bisbas09}
{Bisbas} T.~G.,  {W{\"u}nsch} R.,  {Whitworth} A.~P.,   {Hubber} D.~A.,  2009,
  \mn@doi [\aap] {10.1051/0004-6361/200811522}, \href
  {https://ui.adsabs.harvard.edu/abs/2009A&A...497..649B} {497, 649}

\bibitem[\protect\citeauthoryear{{Bisbas}, {Tanaka}, {Tan}, {Wu}  \&
  {Nakamura}}{{Bisbas} et~al.}{2017}]{bisbas17}
{Bisbas} T.~G.,  {Tanaka} K. E.~I.,  {Tan} J.~C.,  {Wu} B.,   {Nakamura} F.,
  2017, \mn@doi [\apj] {10.3847/1538-4357/aa94c5}, \href
  {https://ui.adsabs.harvard.edu/abs/2017ApJ...850...23B} {850, 23}

\bibitem[\protect\citeauthoryear{{Bock}, {Large}  \& {Sadler}}{{Bock}
  et~al.}{1999}]{bock99}
{Bock} D.~C.~J.,  {Large} M.~I.,   {Sadler} E.~M.,  1999, \mn@doi [\aj]
  {10.1086/300786}, \href
  {https://ui.adsabs.harvard.edu/abs/1999AJ....117.1578B} {117, 1578}

\bibitem[\protect\citeauthoryear{{Boldyrev}}{{Boldyrev}}{2002}]{boldyrev02}
{Boldyrev} S.,  2002, \mn@doi [\apj] {10.1086/339403}, \href
  {https://ui.adsabs.harvard.edu/abs/2002ApJ...569..841B} {569, 841}

\bibitem[\protect\citeauthoryear{{Braiding} et~al.,}{{Braiding}
  et~al.}{2018}]{braiding18}
{Braiding} C.,  et~al., 2018, \mn@doi [\pasa] {10.1017/pasa.2018.18}, \href
  {https://ui.adsabs.harvard.edu/abs/2018PASA...35...29B} {35, e029}

\bibitem[\protect\citeauthoryear{{Bressert}, {Ginsburg}, {Bally}, {Battersby},
  {Longmore}  \& {Testi}}{{Bressert} et~al.}{2012}]{bressert12}
{Bressert} E.,  {Ginsburg} A.,  {Bally} J.,  {Battersby} C.,  {Longmore} S.,
  {Testi} L.,  2012, \mn@doi [\apjl] {10.1088/2041-8205/758/2/L28}, \href
  {https://ui.adsabs.harvard.edu/abs/2012ApJ...758L..28B} {758, L28}

\bibitem[\protect\citeauthoryear{{Carey} et~al.,}{{Carey}
  et~al.}{2005}]{carey05}
{Carey} S.~J.,  et~al., 2005, in American Astronomical Society Meeting
  Abstracts. p. 63.33

\bibitem[\protect\citeauthoryear{{Chira}, {Ib{\'a}{\~n}ez-Mej{\'\i}a}, {Mac
  Low}  \& {Henning}}{{Chira} et~al.}{2019}]{chira19}
{Chira} R.~A.,  {Ib{\'a}{\~n}ez-Mej{\'\i}a} J.~C.,  {Mac Low} M.~M.,
  {Henning} T.,  2019, \mn@doi [\aap] {10.1051/0004-6361/201833970}, \href
  {https://ui.adsabs.harvard.edu/abs/2019A&A...630A..97C} {630, A97}

\bibitem[\protect\citeauthoryear{{Churchwell} et~al.,}{{Churchwell}
  et~al.}{2006}]{churchwell06}
{Churchwell} E.,  et~al., 2006, \mn@doi [\apj] {10.1086/507015}, \href
  {https://ui.adsabs.harvard.edu/abs/2006ApJ...649..759C} {649, 759}

\bibitem[\protect\citeauthoryear{{Cyganowski} et~al.,}{{Cyganowski}
  et~al.}{2008}]{cyganowski08}
{Cyganowski} C.~J.,  et~al., 2008, \mn@doi [\aj]
  {10.1088/0004-6256/136/6/2391}, \href
  {https://ui.adsabs.harvard.edu/abs/2008AJ....136.2391C} {136, 2391}

\bibitem[\protect\citeauthoryear{{Dewangan}}{{Dewangan}}{2022}]{dewangan2022new}
{Dewangan} L.~K.,  2022, arXiv e-prints, \href
  {https://ui.adsabs.harvard.edu/abs/2022arXiv220402127D} {p. arXiv:2204.02127}

\bibitem[\protect\citeauthoryear{{Dewangan} \& {Ojha}}{{Dewangan} \&
  {Ojha}}{2017}]{dewangan17s235}
{Dewangan} L.~K.,  {Ojha} D.~K.,  2017, \mn@doi [\apj]
  {10.3847/1538-4357/aa8e00}, \href
  {https://ui.adsabs.harvard.edu/abs/2017ApJ...849...65D} {849, 65}

\bibitem[\protect\citeauthoryear{{Dewangan}, {Ojha}, {Luna}, {Anandarao},
  {Ninan}, {Mallick}  \& {Mayya}}{{Dewangan} et~al.}{2016a}]{dewangan16}
{Dewangan} L.~K.,  {Ojha} D.~K.,  {Luna} A.,  {Anandarao} B.~G.,  {Ninan}
  J.~P.,  {Mallick} K.~K.,   {Mayya} Y.~D.,  2016a, \mn@doi [\apj]
  {10.3847/0004-637X/819/1/66}, \href
  {https://ui.adsabs.harvard.edu/abs/2016ApJ...819...66D} {819, 66}

\bibitem[\protect\citeauthoryear{{Dewangan}, {Baug}, {Ojha}, {Janardhan},
  {Ninan}, {Luna}  \& {Zinchenko}}{{Dewangan} et~al.}{2016b}]{dewangan16xs}
{Dewangan} L.~K.,  {Baug} T.,  {Ojha} D.~K.,  {Janardhan} P.,  {Ninan} J.~P.,
  {Luna} A.,   {Zinchenko} I.,  2016b, \mn@doi [\apj]
  {10.3847/0004-637X/826/1/27}, \href
  {https://ui.adsabs.harvard.edu/abs/2016ApJ...826...27D} {826, 27}

\bibitem[\protect\citeauthoryear{{Dewangan}, {Ojha}, {Zinchenko}, {Janardhan}
  \& {Luna}}{{Dewangan} et~al.}{2017a}]{dewangan17a}
{Dewangan} L.~K.,  {Ojha} D.~K.,  {Zinchenko} I.,  {Janardhan} P.,   {Luna} A.,
   2017a, \mn@doi [\apj] {10.3847/1538-4357/834/1/22}, \href
  {https://ui.adsabs.harvard.edu/abs/2017ApJ...834...22D} {834, 22}

\bibitem[\protect\citeauthoryear{{Dewangan}, {Ojha}  \& {Baug}}{{Dewangan}
  et~al.}{2017b}]{dewangan17b}
{Dewangan} L.~K.,  {Ojha} D.~K.,   {Baug} T.,  2017b, \mn@doi [\apj]
  {10.3847/1538-4357/aa79a5}, \href
  {https://ui.adsabs.harvard.edu/abs/2017ApJ...844...15D} {844, 15}

\bibitem[\protect\citeauthoryear{{Dewangan}, {Ojha}, {Zinchenko}  \&
  {Baug}}{{Dewangan} et~al.}{2018a}]{dewangan18b}
{Dewangan} L.~K.,  {Ojha} D.~K.,  {Zinchenko} I.,   {Baug} T.,  2018a, \mn@doi
  [\apj] {10.3847/1538-4357/aac6bb}, \href
  {https://ui.adsabs.harvard.edu/abs/2018ApJ...861...19D} {861, 19}

\bibitem[\protect\citeauthoryear{{Dewangan}, {Dhanya}, {Ojha}  \&
  {Zinchenko}}{{Dewangan} et~al.}{2018b}]{dewangan18N36}
{Dewangan} L.~K.,  {Dhanya} J.~S.,  {Ojha} D.~K.,   {Zinchenko} I.,  2018b,
  \mn@doi [\apj] {10.3847/1538-4357/aadfe3}, \href
  {https://ui.adsabs.harvard.edu/abs/2018ApJ...866...20D} {866, 20}

\bibitem[\protect\citeauthoryear{{Dewangan}, {Baug}, {Ojha}  \&
  {Ghosh}}{{Dewangan} et~al.}{2018c}]{dewangan18}
{Dewangan} L.~K.,  {Baug} T.,  {Ojha} D.~K.,   {Ghosh} S.~K.,  2018c, \mn@doi
  [\apj] {10.3847/1538-4357/aae9db}, \href
  {https://ui.adsabs.harvard.edu/abs/2018ApJ...869...30D} {869, 30}

\bibitem[\protect\citeauthoryear{{Dewangan}, {Ojha}, {Baug}  \&
  {Devaraj}}{{Dewangan} et~al.}{2019a}]{dewangan19}
{Dewangan} L.~K.,  {Ojha} D.~K.,  {Baug} T.,   {Devaraj} R.,  2019a, \mn@doi
  [\apj] {10.3847/1538-4357/ab10dc}, \href
  {https://ui.adsabs.harvard.edu/abs/2019ApJ...875..138D} {875, 138}

\bibitem[\protect\citeauthoryear{{Dewangan}, {Pirogov}, {Ryabukhina}, {Ojha}
  \& {Zinchenko}}{{Dewangan} et~al.}{2019b}]{dewangan19x}
{Dewangan} L.~K.,  {Pirogov} L.~E.,  {Ryabukhina} O.~L.,  {Ojha} D.~K.,
  {Zinchenko} I.,  2019b, \mn@doi [\apj] {10.3847/1538-4357/ab1aa6}, \href
  {https://ui.adsabs.harvard.edu/abs/2019ApJ...877....1D} {877, 1}

\bibitem[\protect\citeauthoryear{{Dewangan}, {Dhanya}, {Bhadari}, {Ojha}  \&
  {Baug}}{{Dewangan} et~al.}{2021}]{dewangan21}
{Dewangan} L.~K.,  {Dhanya} J.~S.,  {Bhadari} N.~K.,  {Ojha} D.~K.,   {Baug}
  T.,  2021, \mn@doi [\mnras] {10.1093/mnras/stab2137}, \href
  {https://ui.adsabs.harvard.edu/abs/2021MNRAS.506.6081D} {506, 6081}

\bibitem[\protect\citeauthoryear{{Dhanya}, {Dewangan}, {Ojha}  \&
  {Mandal}}{{Dhanya} et~al.}{2021}]{dhanya21}
{Dhanya} J.~S.,  {Dewangan} L.~K.,  {Ojha} D.~K.,   {Mandal} S.,  2021, \mn@doi
  [\pasj] {10.1093/pasj/psz137}, \href
  {https://ui.adsabs.harvard.edu/abs/2021PASJ...73S.355D} {73, S355}

\bibitem[\protect\citeauthoryear{{Dyson} \& {Williams}}{{Dyson} \&
  {Williams}}{1997}]{dyson97}
{Dyson} J.~E.,  {Williams} D.~A.,  1997, {The physics of the interstellar
  medium}, \mn@doi{10.1201/9780585368115.
}

\bibitem[\protect\citeauthoryear{{Emig} et~al.,}{{Emig} et~al.}{2022}]{emig22}
{Emig} K.~L.,  et~al., 2022, arXiv e-prints, \href
  {https://ui.adsabs.harvard.edu/abs/2022arXiv220509193E} {p. arXiv:2205.09193}

\bibitem[\protect\citeauthoryear{{Enokiya}, {Torii}  \& {Fukui}}{{Enokiya}
  et~al.}{2021}]{Enokiya21}
{Enokiya} R.,  {Torii} K.,   {Fukui} Y.,  2021, \mn@doi [\pasj]
  {10.1093/pasj/psz119}, \href
  {https://ui.adsabs.harvard.edu/abs/2021PASJ...73S..75E} {73, S75}

\bibitem[\protect\citeauthoryear{{Evans} Neal~J. et~al.,}{{Evans}
  et~al.}{2009}]{evans09}
{Evans} Neal~J. I.,  et~al., 2009, \mn@doi [\apjs]
  {10.1088/0067-0049/181/2/321}, \href
  {https://ui.adsabs.harvard.edu/abs/2009ApJS..181..321E} {181, 321}

\bibitem[\protect\citeauthoryear{{Fabricius} et~al.,}{{Fabricius}
  et~al.}{2021}]{fabricius21}
{Fabricius} C.,  et~al., 2021, \mn@doi [\aap] {10.1051/0004-6361/202039834},
  \href {https://ui.adsabs.harvard.edu/abs/2021A&A...649A...5F} {649, A5}

\bibitem[\protect\citeauthoryear{{Foster} et~al.,}{{Foster}
  et~al.}{2011}]{foster11}
{Foster} J.~B.,  et~al., 2011, \mn@doi [\apjs] {10.1088/0067-0049/197/2/25},
  \href {https://ui.adsabs.harvard.edu/abs/2011ApJS..197...25F} {197, 25}

\bibitem[\protect\citeauthoryear{{Fukui} et~al.,}{{Fukui}
  et~al.}{2014}]{fukui14}
{Fukui} Y.,  et~al., 2014, \mn@doi [\apj] {10.1088/0004-637X/780/1/36}, \href
  {https://ui.adsabs.harvard.edu/abs/2014ApJ...780...36F} {780, 36}

\bibitem[\protect\citeauthoryear{{Fukui} et~al.,}{{Fukui}
  et~al.}{2018}]{fukui18}
{Fukui} Y.,  et~al., 2018, \mn@doi [\apj] {10.3847/1538-4357/aac217}, \href
  {https://ui.adsabs.harvard.edu/abs/2018ApJ...859..166F} {859, 166}

\bibitem[\protect\citeauthoryear{{Fukui} et~al.,}{{Fukui}
  et~al.}{2019}]{fukui19ex}
{Fukui} Y.,  et~al., 2019, \mn@doi [\apj] {10.3847/1538-4357/ab4900}, \href
  {https://ui.adsabs.harvard.edu/abs/2019ApJ...886...14F} {886, 14}

\bibitem[\protect\citeauthoryear{{Fukui}, {Habe}, {Inoue}, {Enokiya}  \&
  {Tachihara}}{{Fukui} et~al.}{2021}]{fukui21}
{Fukui} Y.,  {Habe} A.,  {Inoue} T.,  {Enokiya} R.,   {Tachihara} K.,  2021,
  \mn@doi [\pasj] {10.1093/pasj/psaa103}, \href
  {https://ui.adsabs.harvard.edu/abs/2021PASJ...73S...1F} {73, S1}

\bibitem[\protect\citeauthoryear{{Gaczkowski} et~al.,}{{Gaczkowski}
  et~al.}{2015}]{gaczkowski15}
{Gaczkowski} B.,  et~al., 2015, \mn@doi [\aap] {10.1051/0004-6361/201526527},
  \href {https://ui.adsabs.harvard.edu/abs/2015A&A...584A..36G} {584, A36}

\bibitem[\protect\citeauthoryear{{Gaczkowski} et~al.,}{{Gaczkowski}
  et~al.}{2017}]{gaczkowski17}
{Gaczkowski} B.,  et~al., 2017, \mn@doi [\aap] {10.1051/0004-6361/201628508},
  \href {https://ui.adsabs.harvard.edu/abs/2017A&A...608A.102G} {608, A102}

\bibitem[\protect\citeauthoryear{{Gaia Collaboration} et~al.,}{{Gaia
  Collaboration} et~al.}{2021}]{gaia21}
{Gaia Collaboration} et~al., 2021, \mn@doi [\aap]
  {10.1051/0004-6361/202039657}, \href
  {https://ui.adsabs.harvard.edu/abs/2021A&A...649A...1G} {649, A1}

\bibitem[\protect\citeauthoryear{{Garay}, {Brooks}, {Mardones}  \&
  {Norris}}{{Garay} et~al.}{2006}]{garay06}
{Garay} G.,  {Brooks} K.~J.,  {Mardones} D.,   {Norris} R.~P.,  2006, \mn@doi
  [\apj] {10.1086/508048}, \href
  {https://ui.adsabs.harvard.edu/abs/2006ApJ...651..914G} {651, 914}

\bibitem[\protect\citeauthoryear{{Garay}, {Mardones}, {Brooks}, {Videla}  \&
  {Contreras}}{{Garay} et~al.}{2007}]{garay07}
{Garay} G.,  {Mardones} D.,  {Brooks} K.~J.,  {Videla} L.,   {Contreras} Y.,
  2007, \mn@doi [\apj] {10.1086/520103}, \href
  {https://ui.adsabs.harvard.edu/abs/2007ApJ...666..309G} {666, 309}

\bibitem[\protect\citeauthoryear{{Getman}, {Feigelson}, {Garmire}, {Broos}  \&
  {Wang}}{{Getman} et~al.}{2007}]{getman07}
{Getman} K.~V.,  {Feigelson} E.~D.,  {Garmire} G.,  {Broos} P.,   {Wang} J.,
  2007, \mn@doi [\apj] {10.1086/509112}, \href
  {https://ui.adsabs.harvard.edu/abs/2007ApJ...654..316G} {654, 316}

\bibitem[\protect\citeauthoryear{{Gonzalez} \& {Woods}}{{Gonzalez} \&
  {Woods}}{2002}]{gonzalez02}
{Gonzalez} R.~C.,  {Woods} R.~E.,  2002, {Digital image processing}, 2nd edn.
Prentice-Hall, Englewood Cliffs, NJ

\bibitem[\protect\citeauthoryear{{Habe} \& {Ohta}}{{Habe} \&
  {Ohta}}{1992}]{habe92}
{Habe} A.,  {Ohta} K.,  1992, \pasj, \href
  {https://ui.adsabs.harvard.edu/abs/1992PASJ...44..203H} {44, 203}

\bibitem[\protect\citeauthoryear{{Hacar}, {Kainulainen}, {Tafalla}, {Beuther}
  \& {Alves}}{{Hacar} et~al.}{2016}]{hacar16}
{Hacar} A.,  {Kainulainen} J.,  {Tafalla} M.,  {Beuther} H.,   {Alves} J.,
  2016, \mn@doi [\aap] {10.1051/0004-6361/201526015}, \href
  {https://ui.adsabs.harvard.edu/abs/2016A&A...587A..97H} {587, A97}

\bibitem[\protect\citeauthoryear{{Hanaoka} et~al.,}{{Hanaoka}
  et~al.}{2020}]{hanaoka20}
{Hanaoka} M.,  et~al., 2020, \mn@doi [\pasj] {10.1093/pasj/psz123}, \href
  {https://ui.adsabs.harvard.edu/abs/2020PASJ...72....5H} {72, 5}

\bibitem[\protect\citeauthoryear{{Hartmann}, {Megeath}, {Allen}, {Luhman},
  {Calvet}, {D'Alessio}, {Franco-Hernandez}  \& {Fazio}}{{Hartmann}
  et~al.}{2005}]{hartmann05}
{Hartmann} L.,  {Megeath} S.~T.,  {Allen} L.,  {Luhman} K.,  {Calvet} N.,
  {D'Alessio} P.,  {Franco-Hernandez} R.,   {Fazio} G.,  2005, \mn@doi [\apj]
  {10.1086/431472}, \href
  {https://ui.adsabs.harvard.edu/abs/2005ApJ...629..881H} {629, 881}

\bibitem[\protect\citeauthoryear{{Haworth} et~al.,}{{Haworth}
  et~al.}{2015a}]{haworth15a}
{Haworth} T.~J.,  et~al., 2015a, \mn@doi [\mnras] {10.1093/mnras/stv639}, \href
  {https://ui.adsabs.harvard.edu/abs/2015MNRAS.450...10H} {450, 10}

\bibitem[\protect\citeauthoryear{{Haworth}, {Shima}, {Tasker}, {Fukui},
  {Torii}, {Dale}, {Takahira}  \& {Habe}}{{Haworth} et~al.}{2015b}]{haworth15b}
{Haworth} T.~J.,  {Shima} K.,  {Tasker} E.~J.,  {Fukui} Y.,  {Torii} K.,
  {Dale} J.~E.,  {Takahira} K.,   {Habe} A.,  2015b, \mn@doi [\mnras]
  {10.1093/mnras/stv2068}, \href
  {https://ui.adsabs.harvard.edu/abs/2015MNRAS.454.1634H} {454, 1634}

\bibitem[\protect\citeauthoryear{{Henshaw}, {Caselli}, {Fontani},
  {Jim{\'e}nez-Serra}, {Tan}  \& {Hernandez}}{{Henshaw}
  et~al.}{2013}]{henshaw13}
{Henshaw} J.~D.,  {Caselli} P.,  {Fontani} F.,  {Jim{\'e}nez-Serra} I.,  {Tan}
  J.~C.,   {Hernandez} A.~K.,  2013, \mn@doi [\mnras] {10.1093/mnras/sts282},
  \href {https://ui.adsabs.harvard.edu/abs/2013MNRAS.428.3425H} {428, 3425}

\bibitem[\protect\citeauthoryear{{Heyer} \& {Brunt}}{{Heyer} \&
  {Brunt}}{2004}]{heyer04}
{Heyer} M.~H.,  {Brunt} C.~M.,  2004, \mn@doi [\apjl] {10.1086/425978}, \href
  {https://ui.adsabs.harvard.edu/abs/2004ApJ...615L..45H} {615, L45}

\bibitem[\protect\citeauthoryear{{Hirota}}{{Hirota}}{2018}]{hirota18}
{Hirota} T.,  2018, \mn@doi [Publication of Korean Astronomical Society]
  {10.5303/PKAS.2018.33.2.021}, \href
  {https://ui.adsabs.harvard.edu/abs/2018PKAS...33...21H} {33, 21}

\bibitem[\protect\citeauthoryear{{Inoue} \& {Fukui}}{{Inoue} \&
  {Fukui}}{2013}]{inoue13}
{Inoue} T.,  {Fukui} Y.,  2013, \mn@doi [\apjl] {10.1088/2041-8205/774/2/L31},
  \href {https://ui.adsabs.harvard.edu/abs/2013ApJ...774L..31I} {774, L31}

\bibitem[\protect\citeauthoryear{{Inoue}, {Hennebelle}, {Fukui}, {Matsumoto},
  {Iwasaki}  \& {Inutsuka}}{{Inoue} et~al.}{2018}]{inoue18}
{Inoue} T.,  {Hennebelle} P.,  {Fukui} Y.,  {Matsumoto} T.,  {Iwasaki} K.,
  {Inutsuka} S.-i.,  2018, \mn@doi [\pasj] {10.1093/pasj/psx089}, \href
  {https://ui.adsabs.harvard.edu/abs/2018PASJ...70S..53I} {70, S53}

\bibitem[\protect\citeauthoryear{{Jackson} et~al.,}{{Jackson}
  et~al.}{2013}]{jackson13}
{Jackson} J.~M.,  et~al., 2013, \mn@doi [\pasa] {10.1017/pasa.2013.37}, \href
  {https://ui.adsabs.harvard.edu/abs/2013PASA...30...57J} {30, e057}

\bibitem[\protect\citeauthoryear{{Karr} \& {Martin}}{{Karr} \&
  {Martin}}{2003}]{karr03}
{Karr} J.~L.,  {Martin} P.~G.,  2003, \mn@doi [\apj] {10.1086/376895}, \href
  {https://ui.adsabs.harvard.edu/abs/2003ApJ...595..880K} {595, 880}

\bibitem[\protect\citeauthoryear{{Kohno} et~al.,}{{Kohno}
  et~al.}{2018}]{Kohno18}
{Kohno} M.,  et~al., 2018, \mn@doi [\pasj] {10.1093/pasj/psx137}, \href
  {https://ui.adsabs.harvard.edu/abs/2018PASJ...70S..50K} {70, S50}

\bibitem[\protect\citeauthoryear{{Krause} et~al.,}{{Krause}
  et~al.}{2018}]{krause18}
{Krause} M. G.~H.,  et~al., 2018, \mn@doi [\aap] {10.1051/0004-6361/201732416},
  \href {https://ui.adsabs.harvard.edu/abs/2018A&A...619A.120K} {619, A120}

\bibitem[\protect\citeauthoryear{{Kwan}}{{Kwan}}{1997}]{kwan97}
{Kwan} J.,  1997, \mn@doi [\apj] {10.1086/304773}, \href
  {https://ui.adsabs.harvard.edu/abs/1997ApJ...489..284K} {489, 284}

\bibitem[\protect\citeauthoryear{{Lamers} \& {Cassinelli}}{{Lamers} \&
  {Cassinelli}}{1999}]{lamers99}
{Lamers} H. J.~G.~L.~M.,  {Cassinelli} J.~P.,  1999, {Introduction to Stellar
  Winds}

\bibitem[\protect\citeauthoryear{{Larson}}{{Larson}}{1981}]{larson81}
{Larson} R.~B.,  1981, \mn@doi [\mnras] {10.1093/mnras/194.4.809}, \href
  {https://ui.adsabs.harvard.edu/abs/1981MNRAS.194..809L} {194, 809}

\bibitem[\protect\citeauthoryear{{Liang}, {Xu}, {Xu}  \& {Wang}}{{Liang}
  et~al.}{2021}]{liang21}
{Liang} X.,  {Xu} J.-L.,  {Xu} Y.,   {Wang} J.-J.,  2021, \mn@doi [\apj]
  {10.3847/1538-4357/abf1eb}, \href
  {https://ui.adsabs.harvard.edu/abs/2021ApJ...913...14L} {913, 14}

\bibitem[\protect\citeauthoryear{{Marsh}, {Whitworth}  \& {Lomax}}{{Marsh}
  et~al.}{2015}]{marsh15}
{Marsh} K.~A.,  {Whitworth} A.~P.,   {Lomax} O.,  2015, \mn@doi [\mnras]
  {10.1093/mnras/stv2248}, \href
  {https://ui.adsabs.harvard.edu/abs/2015MNRAS.454.4282M} {454, 4282}

\bibitem[\protect\citeauthoryear{{Marsh} et~al.,}{{Marsh}
  et~al.}{2017}]{marsh17}
{Marsh} K.~A.,  et~al., 2017, \mn@doi [\mnras] {10.1093/mnras/stx1723}, \href
  {https://ui.adsabs.harvard.edu/abs/2017MNRAS.471.2730M} {471, 2730}

\bibitem[\protect\citeauthoryear{{Matsakis}, {Evans}, {Sato}  \&
  {Zuckerman}}{{Matsakis} et~al.}{1976}]{matsakis76}
{Matsakis} D.~N.,  {Evans} N.~J. I.,  {Sato} T.,   {Zuckerman} B.,  1976,
  \mn@doi [\aj] {10.1086/111871}, \href
  {https://ui.adsabs.harvard.edu/abs/1976AJ.....81..172M} {81, 172}

\bibitem[\protect\citeauthoryear{{Molinari} et~al.,}{{Molinari}
  et~al.}{2010a}]{Molinari10b}
{Molinari} S.,  et~al., 2010a, \mn@doi [\pasp] {10.1086/651314}, \href
  {https://ui.adsabs.harvard.edu/abs/2010PASP..122..314M} {122, 314}

\bibitem[\protect\citeauthoryear{{Molinari} et~al.,}{{Molinari}
  et~al.}{2010b}]{Molinari10a}
{Molinari} S.,  et~al., 2010b, \mn@doi [\aap] {10.1051/0004-6361/201014659},
  \href {https://ui.adsabs.harvard.edu/abs/2010A&A...518L.100M} {518, L100}

\bibitem[\protect\citeauthoryear{{Morales}, {Mardones}, {Garay}, {Brooks}  \&
  {Pineda}}{{Morales} et~al.}{2009}]{morales09}
{Morales} E. F.~E.,  {Mardones} D.,  {Garay} G.,  {Brooks} K.~J.,   {Pineda}
  J.~E.,  2009, \mn@doi [\apj] {10.1088/0004-637X/698/1/488}, \href
  {https://ui.adsabs.harvard.edu/abs/2009ApJ...698..488M} {698, 488}

\bibitem[\protect\citeauthoryear{{Motte}, {Bontemps}  \& {Louvet}}{{Motte}
  et~al.}{2018}]{Motte+2018}
{Motte} F.,  {Bontemps} S.,   {Louvet} F.,  2018, \mn@doi [\araa]
  {10.1146/annurev-astro-091916-055235}, \href
  {https://ui.adsabs.harvard.edu/abs/2018ARA&A..56...41M} {56, 41}

\bibitem[\protect\citeauthoryear{{Myers}}{{Myers}}{2009}]{myers09}
{Myers} P.~C.,  2009, \mn@doi [\apj] {10.1088/0004-637X/700/2/1609}, \href
  {https://ui.adsabs.harvard.edu/abs/2009ApJ...700.1609M} {700, 1609}

\bibitem[\protect\citeauthoryear{{Panagia}}{{Panagia}}{1973}]{panagia73}
{Panagia} N.,  1973, \mn@doi [\aj] {10.1086/111498}, \href
  {https://ui.adsabs.harvard.edu/abs/1973AJ.....78..929P} {78, 929}

\bibitem[\protect\citeauthoryear{{Pon}, {Johnstone}, {Bally}  \&
  {Heiles}}{{Pon} et~al.}{2014}]{pon14}
{Pon} A.,  {Johnstone} D.,  {Bally} J.,   {Heiles} C.,  2014, \mn@doi [\mnras]
  {10.1093/mnras/stu620}, \href
  {https://ui.adsabs.harvard.edu/abs/2014MNRAS.441.1095P} {441, 1095}

\bibitem[\protect\citeauthoryear{{Priestley} \& {Whitworth}}{{Priestley} \&
  {Whitworth}}{2021}]{Priestley21}
{Priestley} F.~D.,  {Whitworth} A.~P.,  2021, \mn@doi [\mnras]
  {10.1093/mnras/stab1777}, \href
  {https://ui.adsabs.harvard.edu/abs/2021MNRAS.506..775P} {506, 775}

\bibitem[\protect\citeauthoryear{{Rosen}, {Offner}, {Sadavoy}, {Bhandare},
  {V{\'a}zquez-Semadeni}  \& {Ginsburg}}{{Rosen} et~al.}{2020}]{rosen20}
{Rosen} A.~L.,  {Offner} S. S.~R.,  {Sadavoy} S.~I.,  {Bhandare} A.,
  {V{\'a}zquez-Semadeni} E.,   {Ginsburg} A.,  2020, \mn@doi [\ssr]
  {10.1007/s11214-020-00688-5}, \href
  {https://ui.adsabs.harvard.edu/abs/2020SSRv..216...62R} {216, 62}

\bibitem[\protect\citeauthoryear{{Schuller} et~al.,}{{Schuller}
  et~al.}{2009}]{schuller09}
{Schuller} F.,  et~al., 2009, \mn@doi [\aap] {10.1051/0004-6361/200811568},
  \href {https://ui.adsabs.harvard.edu/abs/2009A&A...504..415S} {504, 415}

\bibitem[\protect\citeauthoryear{{Schuller} et~al.,}{{Schuller}
  et~al.}{2017}]{schuller17}
{Schuller} F.,  et~al., 2017, \mn@doi [\aap] {10.1051/0004-6361/201628933},
  \href {https://ui.adsabs.harvard.edu/abs/2017A&A...601A.124S} {601, A124}

\bibitem[\protect\citeauthoryear{{Schuller} et~al.,}{{Schuller}
  et~al.}{2021}]{schuller21}
{Schuller} F.,  et~al., 2021, \mn@doi [\mnras] {10.1093/mnras/staa2369}, \href
  {https://ui.adsabs.harvard.edu/abs/2021MNRAS.500.3064S} {500, 3064}

\bibitem[\protect\citeauthoryear{She \& Leveque}{She \& Leveque}{1994}]{she94}
She Z.-S.,  Leveque E.,  1994, \mn@doi [Phys. Rev. Lett.]
  {10.1103/PhysRevLett.72.336}, 72, 336

\bibitem[\protect\citeauthoryear{{Tan}, {Beltr{\'a}n}, {Caselli}, {Fontani},
  {Fuente}, {Krumholz}, {McKee}  \& {Stolte}}{{Tan} et~al.}{2014}]{tan14}
{Tan} J.~C.,  {Beltr{\'a}n} M.~T.,  {Caselli} P.,  {Fontani} F.,  {Fuente} A.,
  {Krumholz} M.~R.,  {McKee} C.~F.,   {Stolte} A.,  2014, in {Beuther} H.,
  {Klessen} R.~S.,  {Dullemond} C.~P.,   {Henning} T.,  eds, Protostars and
  Planets VI. p.~149 (\mn@eprint {arXiv} {1402.0919}),
  \mn@doi{10.2458/azu\_uapress\_9780816531240-ch007}

\bibitem[\protect\citeauthoryear{{Tig{\'e}} et~al.,}{{Tig{\'e}}
  et~al.}{2017}]{Tige+2017}
{Tig{\'e}} J.,  et~al., 2017, \mn@doi [\aap] {10.1051/0004-6361/201628989},
  \href {https://ui.adsabs.harvard.edu/abs/2017A&A...602A..77T} {602, A77}

\bibitem[\protect\citeauthoryear{{Torii} et~al.,}{{Torii}
  et~al.}{2011}]{torii11}
{Torii} K.,  et~al., 2011, \mn@doi [\apj] {10.1088/0004-637X/738/1/46}, \href
  {https://ui.adsabs.harvard.edu/abs/2011ApJ...738...46T} {738, 46}

\bibitem[\protect\citeauthoryear{{Torii} et~al.,}{{Torii}
  et~al.}{2015}]{torii15}
{Torii} K.,  et~al., 2015, \mn@doi [\apj] {10.1088/0004-637X/806/1/7}, \href
  {https://ui.adsabs.harvard.edu/abs/2015ApJ...806....7T} {806, 7}

\bibitem[\protect\citeauthoryear{{Torii} et~al.,}{{Torii}
  et~al.}{2017}]{torii17}
{Torii} K.,  et~al., 2017, \mn@doi [\apj] {10.3847/1538-4357/835/2/142}, \href
  {https://ui.adsabs.harvard.edu/abs/2017ApJ...835..142T} {835, 142}

\bibitem[\protect\citeauthoryear{{Trevi{\~n}o-Morales}
  et~al.,}{{Trevi{\~n}o-Morales} et~al.}{2019}]{morales19}
{Trevi{\~n}o-Morales} S.~P.,  et~al., 2019, \mn@doi [\aap]
  {10.1051/0004-6361/201935260}, \href
  {https://ui.adsabs.harvard.edu/abs/2019A&A...629A..81T} {629, A81}

\bibitem[\protect\citeauthoryear{{Urquhart} et~al.,}{{Urquhart}
  et~al.}{2018}]{urquhart18}
{Urquhart} J.~S.,  et~al., 2018, \mn@doi [\mnras] {10.1093/mnras/stx2258},
  \href {https://ui.adsabs.harvard.edu/abs/2018MNRAS.473.1059U} {473, 1059}

\bibitem[\protect\citeauthoryear{{Watson}, {Hanspal}  \& {Mengistu}}{{Watson}
  et~al.}{2010}]{watson10}
{Watson} C.,  {Hanspal} U.,   {Mengistu} A.,  2010, \mn@doi [\apj]
  {10.1088/0004-637X/716/2/1478}, \href
  {https://ui.adsabs.harvard.edu/abs/2010ApJ...716.1478W} {716, 1478}

\bibitem[\protect\citeauthoryear{{Whitworth} \& {Priestley}}{{Whitworth} \&
  {Priestley}}{2021}]{whitworth21}
{Whitworth} A.~P.,  {Priestley} F.~D.,  2021, \mn@doi [\mnras]
  {10.1093/mnras/stab1125}, \href
  {https://ui.adsabs.harvard.edu/abs/2021MNRAS.504.3156W} {504, 3156}

\bibitem[\protect\citeauthoryear{{Williams}, {de Geus}  \& {Blitz}}{{Williams}
  et~al.}{1994}]{williams94}
{Williams} J.~P.,  {de Geus} E.~J.,   {Blitz} L.,  1994, \mn@doi [\apj]
  {10.1086/174279}, \href
  {https://ui.adsabs.harvard.edu/abs/1994ApJ...428..693W} {428, 693}

\bibitem[\protect\citeauthoryear{{Wright} et~al.,}{{Wright}
  et~al.}{2010}]{wright10}
{Wright} E.~L.,  et~al., 2010, \mn@doi [\aj] {10.1088/0004-6256/140/6/1868},
  \href {https://ui.adsabs.harvard.edu/abs/2010AJ....140.1868W} {140, 1868}

\bibitem[\protect\citeauthoryear{{Yang} et~al.,}{{Yang} et~al.}{2022}]{yang22}
{Yang} A.~Y.,  et~al., 2022, \mn@doi [\aap] {10.1051/0004-6361/202142039},
  \href {https://ui.adsabs.harvard.edu/abs/2022A&A...658A.160Y} {658, A160}

\makeatother
\end{thebibliography}

%\begin{thebibliography}{99}
%\end{thebibliography}

%%%%%%%%%%%%%%%%%%%%%%%%%%%%%%%%%%%%%%%%%%%%%%%%%%

\end{document}